\documentclass[conference]{IEEEtran}
\usepackage{cite}
\usepackage{amsmath,amssymb,amsfonts}
\usepackage{algorithmic}
\usepackage{graphicx}
\usepackage{textcomp}
\usepackage{xcolor}
\usepackage{tikz}
\usepackage[mode=buildnew]{standalone}
\usepackage{gensymb}
\usepackage{booktabs}
\usetikzlibrary{fit,arrows,positioning,arrows.meta}
\usetikzlibrary{patterns}
\usetikzlibrary{shapes.arrows}
\usepackage{algorithm}
\usepackage{algorithmic}
\usepackage{hyperref}
\usepackage{fancyhdr}
\usepackage{comment}
\usepackage{soul}
\usepackage{subcaption}
\usepackage{pgfplots}
\pgfplotsset{compat=1.18}
\usepgfplotslibrary{groupplots}
\usepackage{enumitem}
\usepackage{balance}
\usepackage{sansmath}
\usetikzlibrary{positioning, shapes.geometric, arrows.meta, shadows, fit, calc, backgrounds}
\pgfdeclarelayer{deepbackground} 
\pgfdeclarelayer{background} 
\pgfdeclarelayer{looplayer}      
\pgfdeclarelayer{modulelayer}    
\pgfsetlayers{deepbackground,background,looplayer,modulelayer,main}

\sffamily
\usepackage{amsmath}
\usepackage{amssymb}
\usepackage{amsfonts}
\usepackage{tikz}

\usetikzlibrary{shapes.geometric, arrows.meta, positioning, fit, calc, backgrounds}

\definecolor{softBlue}{RGB}{232,241,250}
\definecolor{strokeBlue}{RGB}{70,130,180}
\definecolor{softRed}{RGB}{250,235,235}
\definecolor{strokeRed}{RGB}{205,92,92}
\definecolor{softGreen}{RGB}{235,250,235}
\definecolor{strokeGreen}{RGB}{60,179,113}
\definecolor{softOrange}{RGB}{255,248,240}
\definecolor{strokeOrange}{RGB}{255,165,0}
\usepackage{xcolor}
\usetikzlibrary{matrix, positioning}
\usetikzlibrary{shadings}

\definecolor{PRS1}{RGB}{220, 140, 50}   
\definecolor{PRS2}{RGB}{60, 160, 180}   
\definecolor{PRS3}{RGB}{100, 170, 80}   
\definecolor{SIG1}{RGB}{190, 70, 90}    
\definecolor{SIG2}{RGB}{110, 90, 190}   
\definecolor{SIG3}{RGB}{130, 110, 80}   

\newcommand{\sigtxtcol}{black}
\usetikzlibrary{shapes.geometric, positioning, calc, shadows}
\usepackage{pgfplots}
\usetikzlibrary{calc}
\pgfplotsset{compat=1.18}
\usepgfplotslibrary{groupplots}
\pgfplotsset{
    attack zone/.style={
        extra x ticks={301, 900},
        extra x tick labels={},
        extra x tick style={grid=major, major grid style={red, dashed, thick}},
        execute at begin axis={
            \fill[red, opacity=0.15] (axis cs:301, -0.2) rectangle (axis cs:900, 1.2);
        }
    },
    logic plot style/.style={
        width=14cm, height=2.8cm,
        xmin=0, xmax=1200,
        ymin=-0.2, ymax=1.2,
        ytick={0,1},
        yticklabels={},
        yticklabel style={font=\scriptsize\sffamily, align=right}, 
        xtick style={draw=none},
        xticklabels=\empty,
        enlarge x limits=false,
        const plot,
        no markers,
        axis on top,
        axis line style={black!60},
        grid=none
    }
}

\hypersetup{          
    colorlinks=true,
    linkcolor=black,
    filecolor=magenta,      
    urlcolor=cyan,
    citecolor=teal!80,
    pdftitle={},
    }
\def\BibTeX{{\rm B\kern-.05em{\sc i\kern-.025em b}\kern-.08em
    T\kern-.1667em\lower.7ex\hbox{E}\kern-.125emX}}
\begin{document}
\title{Threat Detection and Resilience Techniques in PRS-Assisted OTDOA 5G Positioning Systems}
\author{\IEEEauthorblockN{Thodoris Spanos$^{1,2}$, 
Nikolaos Papageorgiou$^2$, Samuele Fantinato$^3$,\\Nikos Kanistras$^1$, Sergi Due\~nas Pedrosa$^3$, Gianluca Caparra$^4$, Vassilis Paliouras$^2$
\thanks{}
}\vspace{0.1cm}
\IEEEauthorblockA{$^1$Loctio P.C., Greece\\
$^2$Dept. of Electrical and Computer Engineering, University of Patras, Greece\\ 
$^3$Qascom SRL, 36061 Bassano del Grappa, Italy\\
$^4$European Space Research and Technology Center, Noordwijk, The Netherlands}
}
\maketitle
\begin{abstract}
Precise positioning is a key enabler for emerging 5G applications, from autonomous transport to industrial automation. Yet the open physical layer (PL) leaves standard positioning reference signals (PRSs) vulnerable to manipulation. This work addresses the security of downlink observed time difference of arrival positioning (DL-OTDOA) through three contributions. First, we introduce VeriLoc, an open-source system-level simulator designed for realistic channel modeling and PL threat injection. Second, we propose three novel security techniques to enhance resilience and threat detection: encrypted PRS to prevent adversarial waveform synthesis, angular-based source authentication (ABSA), and a cross-layer downlink-uplink handshaking protocol to detect attacks that cannot be mitigated by encryption. Third, utilizing VeriLoc, we evaluate the proposed techniques alongside position tracking and a PRS authentication scheme, which extends the original hash-based message authentication code (HMAC) scheme design to support digital signatures. Simulation results demonstrate that while encryption, authentication schemes, and tracking robustly counter selective PRS spoofing and jamming, the proposed spatial and cross-layer mechanisms are essential for detecting meaconing, collectively maintaining attack detection rates in excess of 90\% while keeping false alarm rates minimal.
\end{abstract}


\section{Introduction}

\noindent Advances in 5G radio access enable high-precision positioning for latency-sensitive and safety-critical applications such as autonomous vehicles, intelligent transportation systems, emergency services, and industrial automation~\cite{italiano, Dwivedi, vehicular, industry}. Consequently, 5G positioning has emerged as a fundamental complement to Global Navigation Satellite Systems (GNSS), enabling hybrid fusion solutions that maintain accuracy even when satellite signals are compromised or unavailable~\cite{Dutta, focarelli2}. 

Downlink observed time difference of arrival (DL‑OTDOA), supported by periodic positioning reference signal (PRS) transmissions, provides a network‑centric multilateration solution in 5G systems, that scales across dense cellular deployments~\cite{italiano, Dwivedi, focarelli2}. However, the openness of the wireless medium, combined with the predictable PRS structure and the lack of native integrity measures at the physical-layer (PL) render the system susceptible to malicious interference~\cite{focarelli2, focarelli3, corsara}. As a result, OTDOA can be undermined by various threats (e.g. spoofing, meaconing, jamming) that degrade accuracy and erode confidence in reported locations.

Unlike GNSS, which has long adopted cryptographic protections such as navigation message authentication (NMA) and spreading code encryption (SCE)~\cite{GNSS}, cellular positioning remains largely protected only at the upper layers through Non-Access Stratum (NAS) and Radio Resource Control (RRC) security~\cite{3gppsec}. However, these protocols secure only the transport of location data after estimation; they do not extend cryptographic protection to the PRS itself, leaving the physical layer domain exposed to signal-level manipulation~\cite{focarelli2}.

Prior work addresses both detection and mitigation of PL threats. Focarelli~\emph{et al.} employ Gaussian mixture models to distinguish legitimate and malicious signals based on channel statistics~\cite{focarelli2}, while Crosara~\emph{et al.} demonstrate spoofing attacks that manipulate auxiliary base station (BS) propagation delays without disrupting the primary communication link~\cite{corsara}. Dutta and Singh emphasize the need for integrated physical-layer and protocol-level security to meet the vertical requirements of 5G, rather than isolated countermeasures~\cite{Dutta}. Addressing the need for secure ranging protocols, Singh~\emph{et al.} introduce V-Range, a scheme that embeds cryptographic integrity checks into the ranging process to verify distance commitments against enlargement attacks~\cite{vrange}. Similarly, Spanos~\emph{et al.} introduce a PRS authentication scheme that embeds hash-based message authentication code (HMAC) into empty PRS resource elements (REs), introducing a similarity threshold to maintain robustness in false alarms under low SNR~\cite{spanos} conditions. Targeting network-side vulnerabilities, Li~\emph{et al.} combine truth discovery clustering with neural networks to identify and trace malicious anchor nodes involved in collusion or jamming attacks~\cite{Li}. Finally, Gao~\emph{et al.} demonstrate that delayed copies of reference signals, superimposed onto specific REs while preserving the communication payload, can exploit the lack of physical layer protection; to mitigate this, they propose identifying transmitters via deep learning-based extraction of unique hardware impairments~\cite{gao1, gao2}.

In this work, we propose a comprehensive PL security framework to enhance the integrity and resilience of PRS‑assisted DL‑OTDOA positioning. We introduce a proactive protection scheme based on the Advanced Encryption Standard in counter mode (AES-CTR) to prevent adversarial PRS synthesis, and extend the PRS authentication method of~\cite{spanos} to support digital signatures (DSs) for explicit source authentication. Furthermore, we address active threats such as meaconing and jamming through angular-based source authentication (ABSA) for spatial verification, a downlink-uplink (DL-UL) handshaking (HS) protocol for position consistency validation across links, and a temporal position tracking algorithm utilizing a Kalman filter (KF) to detect trajectory anomalies.

To validate our approach, we introduce VeriLoc (Verified Location), a MATLAB-based simulator now released as open-source software~\cite{loctio2026veriloc}, which implements PRS-assisted DL-OTDOA positioning under realistic channel propagation. Explicitly designed for PL integrity assessment, VeriLoc supports the generation and injection of adversarial waveforms. Using this framework, we evaluate the robustness of the proposed techniques against these active threats, quantifying detection probabilities and false-alarm rates under representative urban mobility scenarios.

The remainder of this paper is organized as follows. Section~\ref{background} reviews PRS-assisted OTDOA positioning. Section~\ref{sec:system_threats} introduces the VeriLoc framework, describing the system architecture and characterizing the adopted adversarial threat models. Section~\ref{proposed_mechanisms} details the proposed detection and resilience techniques. Section~\ref{simulation_framework} outlines the implementation parameters, while Section~\ref{results} presents the simulation results. Finally, Section~\ref{conclusion} concludes the paper and summarizes the main findings.

\section{PRS and OTDOA Positioning}
\label{background}
\subsection{PRS Signal Structure}
\noindent Positioning in 5G New Radio (NR) relies on the PRS, as specified in~\cite{3gppsrs}. The PRS waveform employs QPSK modulation, where the complex symbols $\mathbf{r}(m)$ are derived from a standardized 31-bit Gold code sequence $\mathbf{c_{P}}(n)$ according to
\begin{equation}
\mathbf{r}(m) = \frac{1 - 2\mathbf{c_{P}}(2m)}{\sqrt{2}} + j \frac{1 - 2\mathbf{c_{P}}(2m+1)}{\sqrt{2}}.
\end{equation}
The underlying pseudo-random sequence generator is initialized with a seed value, $c_{\text{init}}$, which determines the specific sequence used. To ensure sequence orthogonality and manage interference, $c_{\text{init}}$ is defined as a function of the PRS sequence ID ($n_{\text{ID,seq}}^{\text{PRS}}$), the slot number within a frame ($n_{\text{s,f}}^{\mu}$), and the orthogonal frequency division multiplexing (OFDM) symbol index ($l$) within the slot
\begin{align}
\text{c}_{\text{init}} &= \Bigg[ 2^{22} \Bigg\lfloor \frac{n_{\text{ID,seq}}^{\text{PRS}}}{1024} \Bigg\rfloor 
+ 2^{10} \Big(N_\text{symb}^{\text{slot}} n_{\text{s,f}}^{\text{$\mu$}} + l + 1\Big) \nonumber \\
&\quad \times \Big[2 \Big(n_{\text{ID,seq}}^{\text{PRS}} \bmod 1024\Big) + 1\Big] +\notag\\
& \qquad+ \Big(n_{\text{ID,seq}}^{\text{PRS}} \bmod 1024\Big) \Bigg] \bmod 2^{31}.
\end{align}
Because the seed value $c_{\text{init}}$ updates with both the slot number and the symbol index $l$, a distinct Gold sequence is generated for each OFDM symbol. Consequently, since a single PRS resource can span up to twelve symbols in time, up to twelve unique Gold codes must be generated to construct the full PRS instance within a slot.
To further manage inter-cell interference and permit simultaneous PRS transmissions from multiple BSs, 5G NR employs a comb-like frequency structure. Each BS maps its PRS to REs spaced $K_{\mathrm{comb}}^{\text{PRS}} \in \{2, 4, 6, 12\}$ subcarriers apart. Additionally, each BS is assigned a unique frequency offset, $k_{\text{offset}}^{\text{PRS}} \in \{0, 1, \ldots, K_{\mathrm{comb}}^{\text{PRS}} - 1\}$, which shifts its transmission across the frequency grid. This multiplexing enables PRS signals from neighboring BSs to remain orthogonal in the frequency domain, minimizing mutual interference.
\subsection{OTDOA Positioning}\label{Sec_OTDOA}

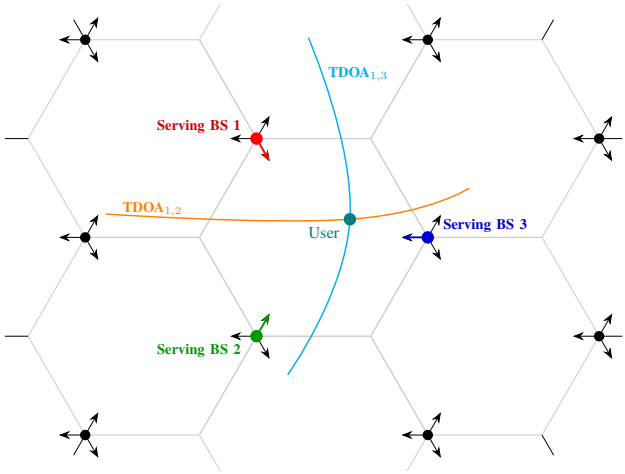
\begin{figure}[htbp]
    \centering
    \resizebox{0.48\textwidth}{!}{%
        \tikzset{
  pics/bsA/.style = {
    code = {
      \draw[-{Stealth[length=2mm,width=1.5mm]}] (0,0) -- (180:0.5cm);
      \draw[-{Stealth[length=2mm,width=1.5mm]}] (0,0) -- (60:0.5cm);
      \draw[-{Stealth[length=2mm,width=1.5mm]}] (0,0) -- (300:0.5cm);
    }
  },
  pics/bsB/.style = {
    code = {
      \draw[-{Stealth[length=2mm,width=1.5mm]}] (0,0) -- (60:0.5cm);
      \draw[-{Stealth[length=2mm,width=1.5mm]}] (0,0) -- (180:0.5cm);
      \draw[-{Stealth[length=2mm,width=1.5mm]}] (0,0) -- (300:0.5cm);
    }
  }
}

\def\R{2.2cm}
\def\V{0.45cm}
\def\D{0.10cm}

\begin{tikzpicture}[every node/.style={inner sep=0, outer sep=0}, transform shape]

\clip (-6, -4.5) rectangle (6, 4.5);

\coordinate (C0) at (0,0);
\coordinate (C1) at ({1.5*\R}, {0.866*\R});
\coordinate (C2) at ({1.5*\R}, {-0.866*\R});
\coordinate (C3) at (0, {-2*0.866*\R});
\coordinate (C4) at ({-1.5*\R}, {-0.866*\R});
\coordinate (C5) at ({-1.5*\R}, {0.866*\R});
\coordinate (C6) at (0, {2*0.866*\R});

\foreach \c in {C0,C1,C2,C3,C4,C5,C6} {
  \draw[black!20] ($(\c)+(0:\R)$) -- ($(\c)+(60:\R)$) -- ($(\c)+(120:\R)$) -- ($(\c)+(180:\R)$)
        -- ($(\c)+(240:\R)$) -- ($(\c)+(300:\R)$) -- cycle;
}

\coordinate (A0)  at ($(C0) + (0:\R)$);
\coordinate (A1)  at ($(C0) + (120:\R)$);
\coordinate (A2)  at ($(C1) + (0:\R)$);
\coordinate (A3)  at ($(C1) + (120:\R)$);
\coordinate (A4)  at ($(C4) + (240:\R)$);
\coordinate (A5)  at ($(C3) + (240:\R)$);
\coordinate (A6)  at ($(C5) + (240:\R)$);
\coordinate (A7)  at ($(C5) + (120:\R)$);
\coordinate (A8)  at ($(C6) + (120:\R)$);
\coordinate (A9)  at ($(C2) + (0:\R)$);
\coordinate (A10) at ($(C3) + (0:\R)$);
\coordinate (A11) at ($(C4) + (0:\R)$);

\coordinate (UserLoc) at (0.7, 0.35);

\draw[orange, thick] plot [smooth, tension=0.8] coordinates {
    ($(UserLoc) + (-4.7, 0.1)$)
    (UserLoc)
    ($(UserLoc) + (2.3, 0.6)$)
};
\node[orange, font=\footnotesize\bfseries, fill=none, inner sep=1pt] at ($(UserLoc) + (-3.8, 0.2)$) {TDOA$_{1,2}$};

\draw[cyan, thick] plot [smooth, tension=0.8] coordinates {
    ($(UserLoc) + (-0.8, 3.5)$)
    (UserLoc)
    ($(UserLoc) + (-1.2, -3.0)$)
};
\node[cyan, font=\footnotesize\bfseries, fill=none, inner sep=1pt, anchor=west] at ($(UserLoc) + (-0.45, 2.8)$) {TDOA$_{1,3}$};

\pic at (A0)  {bsA}; \pic at (A1)  {bsA}; \pic at (A2)  {bsB};
\pic at (A3)  {bsB}; \pic at (A4)  {bsB}; \pic at (A5)  {bsB};
\pic at (A6)  {bsB}; \pic at (A7)  {bsB}; \pic at (A8)  {bsB};
\pic at (A9)  {bsB}; \pic at (A10) {bsA}; \pic at (A11) {bsA};

\foreach \p in {A0,A1,A2,A3,A4,A5,A6,A7,A8,A9,A10,A11} {
  \fill[black] (\p) circle (\D);
}

\newcommand{\vertexTick}[2]{\draw ($ (#1) + (#2:\R) $) -- ++(#2:\V);}
\vertexTick{C1}{0} \vertexTick{C1}{60} \vertexTick{C6}{60}
\vertexTick{C6}{120} \vertexTick{C5}{120} \vertexTick{C5}{180}
\vertexTick{C4}{180} \vertexTick{C4}{240} \vertexTick{C3}{240}
\vertexTick{C3}{300} \vertexTick{C2}{300} \vertexTick{C2} {0}

\fill[teal] (UserLoc) circle (0.12cm);
\node[teal, anchor=east, font=\small, fill=white, inner sep=1.5pt] at ($(UserLoc) + (-0.15,-0.25)$) {User};

\begin{scope}[red,line width=1pt]
  \draw[-{Stealth[length=2mm,width=1.5mm]}] (A1) -- ($ (A1) + (300:0.5cm) $);
  \fill[red] (A1) circle (\D+0.02cm);
\end{scope}
\node[anchor=south east, red!80!black, align=right, font=\bfseries\footnotesize] at ($ (A1) + (-0.3,0.1) $) {Serving BS 1};

\begin{scope}[green!60!black, line width=1pt]
  \draw[-{Stealth[length=2mm,width=1.5mm]}] (A11) -- ($ (A11) + (60:0.5cm) $);
  \fill[green!60!black] (A11) circle (\D+0.02cm);
\end{scope}
\node[anchor=north east, green!60!black, align=right, font=\bfseries\footnotesize] at ($ (A11) + (-0.3,-0.15) $) {Serving BS 2};

\begin{scope}[blue!80!black, line width=1pt]
  \draw[-{Stealth[length=2mm,width=1.5mm]}] (A0) -- ($ (A0) + (180:0.5cm) $);
  \fill[blue!80!black] (A0) circle (\D+0.02cm);
\end{scope}
\node[anchor=south west, blue!80!black, align=left, font=\bfseries\footnotesize, fill=white, opacity=0.9, text opacity=1] at ($ (A0) + (0.3,0.1) $) {Serving BS 3};

\end{tikzpicture}%
    }
    \caption{Intersection of two hyperbolas for a three-Serving BS hexagonal cellular network topology.}
    \label{hyperbola}
\end{figure}

\noindent OTDOA positioning~\cite{otdoa2, 3gppotdoa} estimates the user equipment (UE) position via multilateration of time differences. The UE determines the time of arrival (ToA) for each BS $i$ by finding the sample offset $\hat{t}_i$ that maximizes the cross-correlation between the received signal $\mathbf{\tilde{y}}$ and the local replica $\mathbf{s}_i$
\begin{equation}\label{crosscorrelation}
\hat{t}_i = \operatorname*{argmax}\nolimits_{t} \bigg| \sum\nolimits_{k=1}^{L_{\text{seq}} -1} \mathbf{s}_i^*(k) \mathbf{\tilde{y}}(t+k) \bigg|,
\end{equation}
where $L_{\text{seq}}$ denotes the waveform length in samples and $^*$ represents the complex conjugate. Using the ToA estimate from a reference BS $j$, the reference signal timing difference (RSTD) is computed as 
\begin{equation}
    \Delta \tau_{ij} = \frac{\hat{t}_i - \hat{t}_j}{f_s},
\end{equation}
 where $f_s$ is the sampling rate. Geometrically, the condition 
 \begin{equation}
     c \cdot \Delta \tau_{ij} = \|\mathbf{p}_{\text{UE}} - \mathbf{p}_i\| - \|\mathbf{p}_{\text{UE}} - \mathbf{p}_j\|
 \end{equation}
places the UE on a hyperbola with the two BSs as foci. With a set of $N_{\text{BS}}$ serving BSs, the measured RSTDs generate $N_{\text{BS}}-1$ hyperbolas relative to the reference node. The intersection of these curves yields the estimated UE position $\widehat{\mathbf{p}}_{\text{UE}}$. In an overdetermined system ($N_{\text{BS}} > 3$), multiple intersection points may be averaged or processed to improve accuracy. Fig.~\ref{hyperbola} illustrates the fundamental process for a configuration of $N_{\text{BS}}=3$ BSs, where BS$_1$ serves as the reference, producing two hyperbolas that intersect at the UE location.

\section{VeriLoc and Threat Models}
\label{sec:system_threats}

\noindent This study utilizes the VeriLoc framework to quantify the performance of the security techniques discussed in Section~\ref{proposed_mechanisms}. 

\subsection{VeriLoc System Architecture}

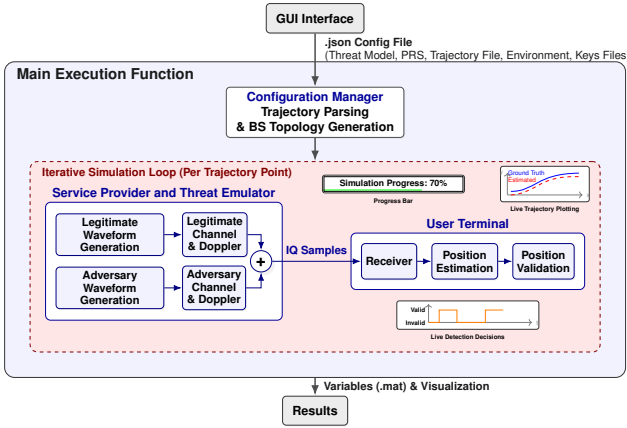
\begin{figure}[htbp]
    \centering
    \resizebox{0.48\textwidth}{!}{%
        \pgfdeclarelayer{deepbackground} 
\pgfdeclarelayer{looplayer}      
\pgfdeclarelayer{modulelayer}    
\pgfsetlayers{deepbackground,looplayer,modulelayer,main}

\sffamily

\begin{tikzpicture}[
    basebox/.style={
        rectangle,
        draw=black!70, thick,
        rounded corners=3pt,
        align=center,
        font=\bfseries\normalsize,
        drop shadow={opacity=0.15, shadow xshift=1pt, shadow yshift=-1pt}
    },
    interfacebox/.style={
        basebox,
        minimum height=0.8cm,
        fill=gray!15, 
        draw=black!80, 
        inner sep=8pt 
    },
    subfuncbox/.style={
        basebox,
        minimum width=3.0cm, minimum height=0.9cm, 
        fill=white!95!blue, 
        draw=blue!50!black,
        font=\bfseries\small,
        align=center
    },
    modulecontainer/.style={
        rectangle,
        draw=blue!60!black, thick,
        fill=white,
        rounded corners=4pt,
        inner sep=8pt
    },
    loopcontainer/.style={
        rectangle,
        draw=red!50!black, thick, dashed,
        fill=red!10!white,
        rounded corners=6pt,
        inner sep=12pt,
        inner ysep=18pt
    },
    maincontainer/.style={
        rectangle,
        draw=black!80, thick, 
        fill=blue!5, 
        rounded corners=8pt,
        inner sep=20pt
    },
    flow/.style={
        ->, >=Latex, very thick, color=black!70
    },
    internalflow/.style={
        ->, >=Latex, thick, color=blue!50!black
    },
    arrowlabel/.style={
        midway, right=3pt, font=\small, align=left, color=black!80
    },
    sumnode/.style={
        circle,
        draw=blue!50!black, thick,
        fill=white,
        inner sep=0pt,
        minimum size=6mm, 
        font=\Large\bfseries, 
        drop shadow={opacity=0.1}
    },
    vizbox/.style={
        rectangle,
        draw=black!80, thin,
        fill=white,
        rounded corners=2pt,
        inner sep=4pt,
        font=\scriptsize\bfseries
    }
]

    \node[subfuncbox, text width=2.8cm] (tx_sig) {Legitimate Waveform\\Generation};
    \node[subfuncbox, text width=2.8cm, below=0.3cm of tx_sig] (tx_threat) {Adversary Waveform\\Generation};
    
    \node[subfuncbox, right=0.5cm of tx_sig, minimum height=0.9cm, minimum width=0.2cm] (tx_chan_legit) {Legitimate\\Channel\\\& Doppler};
    \node[subfuncbox, right=0.5cm of tx_threat, minimum height=0.9cm, minimum width=0.2cm] (tx_chan_adv) {Adversary\\Channel\\\& Doppler};

    \coordinate (chan_mid) at ($(tx_chan_legit.south)!0.5!(tx_chan_adv.north)$);
    
    \node[sumnode, right=1cm of chan_mid] (combiner) {+};

    \draw[internalflow] (tx_sig) -- (tx_chan_legit);
    \draw[internalflow] (tx_threat) -- (tx_chan_adv);
    \draw[internalflow] (tx_chan_legit.east) -| (combiner.north);
    \draw[internalflow] (tx_chan_adv.east) -| (combiner.south);

    \begin{pgfonlayer}{modulelayer}
        \node[modulecontainer, fit=(tx_sig) (tx_threat) (tx_chan_legit) (tx_chan_adv) (combiner)] (tx) {};
        \node[anchor=south, font=\bfseries\normalsize, color=blue!60!black] at (tx.north) {Service Provider and Threat Emulator};
    \end{pgfonlayer}

    \node[subfuncbox, right=2.5cm of combiner, minimum height=1cm, minimum width=1cm] (rx_corr) {Receiver};
    \node[subfuncbox, right=0.4cm of rx_corr, minimum height=1cm, minimum width=1cm] (rx_est) {Position\\Estimation};
    \node[subfuncbox, right=0.4cm of rx_est, minimum height=1cm, minimum width=1cm] (rx_valid) {Position\\Validation};

    \draw[internalflow] (rx_corr) -- (rx_est);
    \draw[internalflow] (rx_est) -- (rx_valid);

    \begin{pgfonlayer}{modulelayer}
        \node[modulecontainer, fit=(rx_corr) (rx_est) (rx_valid)] (rx) {};
        \node[anchor=south, font=\bfseries\normalsize, color=blue!60!black] at (rx.north) {User Terminal};
    \end{pgfonlayer}

    \coordinate (simulation_center) at ($(tx.west)!0.5!(rx.east)$);

    \node[basebox, minimum width=5cm, fill=white, above=1.8cm of simulation_center|-tx.north] (traj) {\textcolor{blue!60!black}{\textbf{Configuration Manager}}\\Trajectory Parsing\\\& BS Topology Generation};

    \node[interfacebox, above=1.5cm of traj] (gui) {GUI Interface};

    \coordinate[below=1.3cm of traj] (loop_top);

    \node[vizbox, right=0.2cm of loop_top, anchor=west, minimum width=4cm, minimum height=0.5cm] (prog_bar) {};
    \fill[green!50] ($(prog_bar.south west)+(0.05,0.05)$) rectangle ($(prog_bar.north west)!0.7!(prog_bar.south east)-(0,0.05)$);
    \draw[black, thin] ($(prog_bar.south west)+(0.05,0.05)$) rectangle ($(prog_bar.north east)+(-0.05,-0.05)$);
    \node[font=\scriptsize\bfseries] at (prog_bar.center) {Simulation Progress: 70\%};
    \node[below=1pt, font=\tiny\bfseries] at (prog_bar.south) {Progress Bar};

    \node[vizbox, right=1cm of prog_bar, minimum width=2.5cm, minimum height=1cm] (live_plot) {};
    
    \draw[->, thin, gray] ($(live_plot.south west)+(0.2,0.2)$) -- ($(live_plot.south east)+(-0.2,0.2)$) node[right, font=\tiny] {x};
    \draw[->, thin, gray] ($(live_plot.south west)+(0.2,0.2)$) -- ($(live_plot.north west)+(0.2,-0.2)$) node[left, font=\tiny] {y};
    
    \draw[blue, thick] ($(live_plot.south west)+(0.3,0.3)$) 
        .. controls ($(live_plot.south west)+(1.0,0.3)$) and ($(live_plot.south west)+(1.0,0.8)$) .. 
        ($(live_plot.north east)+(-0.3,-0.2)$);
        
    \draw[red, dashed, thick] ($(live_plot.south west)+(0.3,0.2)$) 
        .. controls ($(live_plot.south west)+(1.05,0.25)$) and ($(live_plot.south west)+(1.05,0.75)$) .. 
        ($(live_plot.north east)+(-0.3,-0.3)$);
    
    \node[anchor=north west, font=\tiny] at ($(live_plot.north west)+(0.1, 0)$) {\textcolor{blue}{Ground Truth}};
    \node[anchor=north west, font=\tiny] at ($(live_plot.north west)+(0.1, -0.2)$) {\textcolor{red}{Estimated}};
    \node[below, font=\tiny\bfseries] at (live_plot.south) {Live Trajectory Plotting};

    \node[vizbox, below=0.3cm of rx, minimum width=4cm, minimum height=0.8cm] (dec_plot) {};
    
    \draw[->, thin, gray] ($(dec_plot.south west)+(0.9,0.2)$) -- ($(dec_plot.south east)+(-0.2,0.2)$) node[right, font=\tiny] {t};
    \draw[->, thin, gray] ($(dec_plot.south west)+(0.9,0.2)$) -- ($(dec_plot.north west)+(0.9,-0.1)$);

    \node[anchor=east, font=\tiny\bfseries, inner sep=1pt] at ($(dec_plot.south west)+(0.85, 0.2)$) {Invalid};
    \node[anchor=east, font=\tiny\bfseries, inner sep=1pt] at ($(dec_plot.south west)+(0.85, 0.55)$) {Valid};

    \draw[orange, thick] ($(dec_plot.south west)+(0.9,0.2)$) -- ++(0.3,0) -- ++(0,0.35) -- ++(0.5,0) -- ++(0,-0.35) -- ++(0.8,0) -- ++(0,0.35) -- ++(0.5,0);
    
    \node[below, font=\tiny\bfseries] at (dec_plot.south) {Live Detection Decisions};
    
    \begin{pgfonlayer}{looplayer}
        \node[loopcontainer, fit=(tx) (rx) (loop_top) (dec_plot)] (loop) {};
        \node[anchor=north west, font=\bfseries\small, color=red!50!black, inner xsep=10pt, inner ysep=5pt] 
            at (loop.north west) {Iterative Simulation Loop (Per Trajectory Point)};
    \end{pgfonlayer}

    \begin{pgfonlayer}{deepbackground}
        \node[maincontainer, fit=(traj) (loop)] (main) {};
        \node[anchor=north west, font=\bfseries\large, color=black!90, inner xsep=10pt, inner ysep=5pt] 
            at (main.north west) {Main Execution Function};
    \end{pgfonlayer}

    \node[interfacebox, below=0.5cm of main] (results) {Results};

    \draw[flow] (gui.south) -- (traj.north) 
        node[pos=0.3, right=4pt, font=\small, align=left, color=black!90] 
        {\textbf{.json Config File}\\(Threat Model, PRS, Trajectory File, Environment, Keys Files)};

    \draw[flow] (traj.south) -- (loop.north);

    \draw[internalflow] (combiner.east) -- (rx_corr.west) 
        node[midway, above, font=\small\bfseries, color=blue!60!black] {IQ Samples};

    \draw[flow] (main.south) -- (results.north) 
        node[midway, right=3pt, font=\small, align=left, color=black!90] 
        {\textbf{Variables (.mat) \& Visualization}};

\end{tikzpicture}
    }
    \caption{VeriLoc system architecture.}
    \label{software}
\end{figure}
\noindent The high-level architecture of VeriLoc, illustrated in Fig.~\ref{software}, is structurally divided into a graphical user interface (GUI) and a main execution function. The GUI acts as the front-end, enabling users to aggregate configurable simulation parameters, i.e., threat model selection, PRS configuration, cryptographic keys, threat detection techniques, and the trajectory, into a JSON configuration file. Furthermore, the GUI facilitates real-time monitoring by providing progress tracking, dynamic trajectory visualization, and the display of binary validation decisions based on active threat detection techniques.

The main execution function serves as the core simulation engine and comprises three primary modules: a configuration manager, a service provider and threat emulator, and a user terminal. The configuration manager translates the user inputs into the active simulation environment, handling tasks such as parsing the trajectory data and generating the BS topology. The service provider and threat emulator handles the PL simulation by synthesizing both legitimate and adversarial waveforms, propagating them through independent channels with distinct Doppler shifts and path loss effects, and superimposing them into a single stream of composite IQ samples. Finally, the user terminal represents the receiver, which processes this composite signal to estimate the user position and evaluates the integrity of that estimate based on the security techniques enabled via the JSON configuration.


\subsection{VeriLoc System Operational Behavior}
\label{subsec:system_arch}
\begin{figure*}[htbp]
    \centering
    \resizebox{\textwidth}{!}{%
        \begin{tikzpicture}[
    node distance=1.8cm and 0.8cm, 
    auto,
    font=\sffamily\Large, 
    base_block/.style={
        rectangle, 
        rounded corners=3pt, 
        align=center,        
        inner sep=0.8em,     
        minimum height=4em,
        line width=1pt
    },
    block/.style={base_block, draw=strokeBlue, fill=softBlue},
    sec_block/.style={base_block, draw=strokeGreen, fill=softGreen},
    att_block/.style={base_block, draw=strokeRed, fill=softRed},
    ue_block/.style={base_block, draw=strokeOrange, fill=softOrange},
    sum/.style={
        circle, 
        draw=black!70, 
        fill=white, 
        minimum size=3em, 
        inner sep=0pt,
        font=\huge, 
        line width=1pt
    },
    decision/.style={
        diamond, 
        draw=strokeGreen, 
        fill=softGreen, 
        align=center, 
        inner sep=0.2em,     
        aspect=1.5,
        minimum height=4em
    },
    line/.style={
        draw, 
        -{Latex[length=4mm,width=4mm]}, 
        color=black!85, 
        line width=1.2pt
    },
    group_box/.style={
        draw=black!50, 
        dashed, 
        inner xsep=1.5em,
        inner ysep=0.8em,
        rounded corners
    },
    outer_box/.style={
        draw=black!85, 
        dashed, 
        inner xsep=0.4em,
        inner ysep=0.8em, 
        rounded corners,
        line width=1.5pt
    }
]

    \node [block] (prs) {PRS Generation};
    \node [sec_block] (crypto) [right=of prs] {Optional Security:\\1. PRS Encryption\\2. Signature Embedding};
    \node [block] (channel) [right=of crypto] {Channel \& Power Model:\\1. UMa Path Loss\\2. Doppler Shift\\3. Power ($\le$ 24 dBm)};
    
    \path [line] (prs) -- (crypto);
    \path [line] (crypto) -- (channel);

    \node [sum] (sum_all) [right=of channel, xshift=1cm] {$\sum$};
    \node [block, fill=gray!10, draw=gray!60] (awgn) [below=of sum_all, yshift=0cm] {AWGN};

    \path [line] (channel.east) -- node[above, pos=0.7, yshift=0.3em] {$\mathbf{s}_i(t)$} (sum_all.west);
    \path [line] (awgn.north) -- node[left, pos=0.5] {$\mathbf{n}(t)$} (sum_all.south);

    \node [ue_block] (receiver) [right=of sum_all, xshift=0.8cm] {UE RF Front-End};
    \node [ue_block] (doppler) [right=of receiver] {Coarse Doppler\\ Correction};
    \node [ue_block] (rstd) [right=of doppler] {Timing Sync. \&\\ RSTD Correlation};
    \node [ue_block] (rfo) [right=of rstd] {Residual Frequency\\ Offset Compensation};

    \path [line] (sum_all.east) -- node[above, pos=0.35] {$\mathbf{\tilde{y}}(t)$} (receiver.west);
    \path [line] (receiver) -- (doppler);
    \path [line] (doppler) -- (rstd);
    \path [line] (rstd.east) -- (rfo.west);

    \node [sec_block] (extract) [above=1.5cm of rfo] {Signature \&\\ Pilot Extraction};
    \node [decision] (validate) [right=of extract] {Security\\ Validation};
    
    \path [line] (rfo.north) -- (extract.south);
    \path [line] (extract) -- (validate);

    \node [ue_block] (otdoa) [below=0.8cm of rstd] {OTDOA Position Solver};
    \node [ue_block] (pos_out) [right=of otdoa] {Initial Position $\widehat{\mathbf{p}}_{\text{UE}}$};
    
    \path [line] (rstd.south) -- (otdoa.north);
    \path [line] (otdoa) -- (pos_out);

    \node [ue_block, fill=strokeGreen!20, line width=1.5pt] (final_pos) at (validate |- rfo) {Validated\\ Position};

    \path [line] (validate.south) -- node[right, pos=0.4, xshift=0.2cm] {Validity} (final_pos.north);
    \path [line] (pos_out.east) -| (final_pos.south);

    \path let \p1=(prs.west), \p2=(channel.east) in coordinate (tx_center) at ({(\x1+\x2)/2}, 0);
    \coordinate (att_center) at ($(tx_center |- extract.center) + (0, 0.1cm)$); 

    \node [att_block, anchor=east] (attacker) at ([xshift=-0.9cm]att_center) {Attacker Signal Generation};
    \node [att_block, anchor=west] (att_power) at ([xshift=0.9cm]att_center) {Attacker Power Scaling\\($P_{Att} > P_{BS_i}$)};
    
    \path [line] (attacker) -- (att_power);
    \path [line] (att_power.east) -| node[left, pos=0.85, yshift=1em] {$\mathbf{s}_{att}(t)$} (sum_all.north);

    \begin{scope}[on background layer]
        \node [group_box, fit=(prs) (crypto) (channel)] (tx_box) {};
        \node [below, font=\bfseries\Large] (tx_label) at (tx_box.south) {Transmitter Chain (per BS)};
        
        \node [group_box, fit=(attacker) (att_power)] (att_box) {};
        \node [above, font=\bfseries\Large] (att_label) at (att_box.north) {Adversary};
        
        \node[inner sep=0pt, fit=(att_label) (extract) (validate)] (global_top) {};
        \node[inner sep=0pt, fit=(tx_label) (otdoa) (pos_out)] (global_bot) {};

        \node [outer_box, fit=(tx_box) (tx_label) (att_box) (att_label) 
               (tx_box.west |- global_top.north)
               (tx_box.east |- global_bot.south)] (sp_box) {};
               
        \node [below, font=\bfseries\LARGE, yshift=-0.1cm] at (sp_box.south) {Service Provider and Threat Emulator};
        
        \node [group_box, fit=(receiver) (doppler) (rstd) (rfo) (extract) (validate) (otdoa) (pos_out) (final_pos) 
               (receiver.west |- global_top.north)
               (final_pos.east |- global_bot.south)] (rx_box) {};
               
        \node [below, font=\bfseries\LARGE, yshift=-0.1cm] at (rx_box.south) {User Terminal};
    \end{scope}

\end{tikzpicture}%
    }
    \caption{Block diagram of the end-to-end VeriLoc system model: transmitter chain, adversary, channel, and receiver processing.}
    \label{system}
\end{figure*}
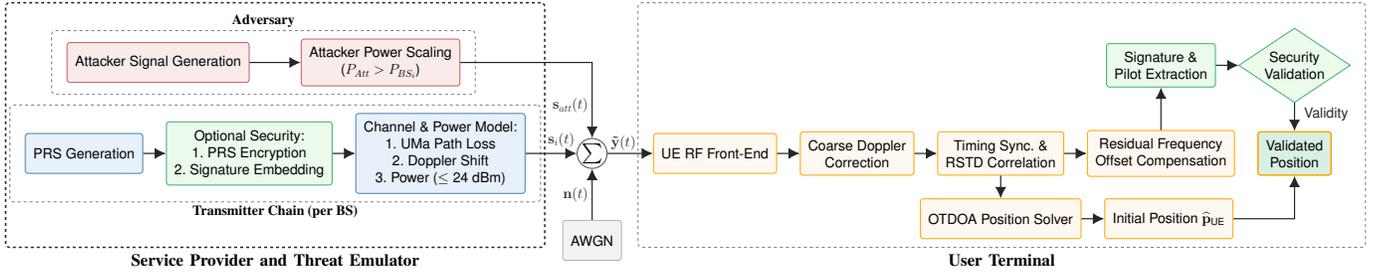
\noindent The end-to-end operational flow of VeriLoc is depicted in Fig.~\ref{system}, encompassing signal generation, channel propagation, adversarial injection, and receiver processing. The system adopts the hexagonal network topology of Fig.~\ref{hyperbola}, where each cell is served by three BSs located at its vertices to satisfy OTDOA trilateration requirements. As the UE traverses cell boundaries, the serving BSs are dynamically updated to maintain continuous positioning coverage.

The transmitter chain configures PL parameters such as carrier frequency, bandwidth, numerology ($\mu$), and PRS comb size ($K_{\mathrm{comb}}^{\mathrm{PRS}}$). Unique PRS sequences are generated for each serving BS based on their ID and mapped onto the radio frame. Additionally, the transmitter integrates a security module capable of applying PRS encryption and authentication, providing the architectural foundation for the corresponding security mechanisms proposed in Section~\ref{proposed_mechanisms}.

Signals propagate through a 3GPP urban macro (UMa) channel, incorporating Doppler shifts relative to the UE velocity. VeriLoc employs idealized power control, scaling transmission power based on channel estimates to ensure equal received power from all serving nodes, bounded at 24~dBm. The received composite signal is a superposition of these legitimate transmissions, additive white Gaussian noise (AWGN), and an adversarial waveform (defined in Section~\ref{subsec:threat_models}). The adversarial signal undergoes identical propagation and Doppler effects but is transmitted at 48~dBm to effectively overshadow legitimate signals. 

To process this composite signal, VeriLoc employs local PRS replicas. Frequency synchronization is performed via a coarse Doppler cross-correlation scan, refined by parabolic interpolation. Following compensation, timing estimation extracts RSTD values via cross-correlation, with residual phase offsets corrected using the complex peak angle. These measurements feed a multilateration solver to estimate the UE position. Finally, the threat detection logic detailed in Section~\ref{proposed_mechanisms} evaluates this estimate, validating or rejecting it based on the decisions of the enabled threat detections techniques.

\subsection{Adversarial Threat Models}
\label{subsec:threat_models}

\noindent The integrity of PRS-assisted OTDOA positioning is based on the assumption that signals arriving at the receiver have not been subjected to adversarial manipulation. However, this assumption fails in environments where adversaries can exploit the signal structure to induce errors in positioning services. VeriLoc models three representative PL threats: false base station (FBS) overshadowing for selective PRS spoofing, meaconing, and jamming, detailed in the following. 
\subsubsection{Selective PRS Spoofing}
\noindent In this attack, an adversary generates a high-power replica PRS waveform using known target BS parameters (e.g., $n_{\text{ID,seq}}^{\text{PRS}}$, $k_{\text{offset}}^{\text{PRS}}$)~\cite{focarelli3, spanos}. Unlike network-level approaches~\cite{10550127, 9509449, 3gppfbs}, this threat model operates exclusively at the PL. 
By applying a specific timing delay, the attacker overshadows the legitimate signal, forcing the UE sample offset estimator in~\eqref{crosscorrelation} to lock onto the malicious rather than the legitimate signal peak.
This introduces a deterministic timing bias to the RSTD measurement~\cite{focarelli2, focarelli3}. When combined with measurements from other BSs, the distorted hyperbolic intersection forces the position estimate $\widehat{\mathbf{p}}_{\text{UE}}$ to converge on false coordinates explicitly chosen by the adversary. Alternatively, if the altered RSTDs are geometrically inconsistent, the multilateration solver fails to converge, resulting in a denial-of-service (DoS).

\subsubsection{Meaconing}
\noindent Meaconing involves intercepting the composite signal from legitimate serving BSs and retransmitting it with amplification, without synthesizing new waveforms~\cite{11027039}. The adversary, positioned at $\widehat{\mathbf{p}}_{\text{Adv}}$, effectively relays the signal geometry at their location to the victim.
Since the adversary amplifies the entire signal uniformly, the RSTDs between BSs are preserved. The common processing delay $\delta$ introduced by the retransmission cancels out in the RSTD computation
\begin{equation}
\Delta\tau_{ij}^{\text{UE}} = (t_i + \delta) - (t_j + \delta) = t_i - t_j = \Delta\tau_{ij}^{\text{Adv}}.
\end{equation}
Consequently, the UE observes the RSTD geometry at the location of the adversary. This effectively projects the location of the malicious node onto the legitimate UE, spoofing the user into believing they are co-located with the meaconing device.

\subsubsection{Jamming Attack}
Unlike deception attacks, jamming aims to cause DoS by transmitting high-power wideband noise across the PRS bandwidth~\cite{arjoune2020smart}. This noise, $\mathbf{n_{jam}}(t)$, significantly exceeds the thermal noise floor.
Unlike spoofing or meaconing, which aim to deceive the receiver into resolving an incorrect location, a jamming attack seeks to degrade the reliability of positioning to the point of DoS~\cite{arjoune2020smart}. We consider an adversary transmitting high-power wideband noise that covers the entire PRS bandwidth.

In the context of OTDOA, BS detection relies on identifying distinct cross-correlation peaks. Jamming injects additive noise $\mathbf{n_{jam}}(t)$ significantly above the thermal floor. When the signal-to-interference-plus-noise ratio (SINR) drops below the minimum sensitivity threshold of the receiver, the correlation peak becomes indistinguishable from the noise floor. Therefore, if an insufficient number of BSs (i.e., $N_{BS} < 3$) are detected, the UE cannot form adequate RSTDs, rendering position calculation impossible.


\section{Proposed Detection and Resilience Techniques}
\label{proposed_mechanisms}

\noindent To counter the threats defined in Section~\ref{subsec:threat_models}, we propose a framework of five techniques that span the cryptographic, signal, spatial, and temporal domains. Although these techniques together provide a multi-layered defense, here we assess each one separately to measure its individual security benefits and trade-offs against specific attack vectors, demonstrating that every technique protects a distinct dimension and collectively delivers robust, detection-based security for OTDOA.

\subsection{PRS Encryption}

\noindent 
To counter a false PRS transmission by an attacker,
we propose encrypting the PRS sequence using AES-CTR, which functions as a synchronous stream cipher and transforms the PRS into a cryptographically secure sequence. Let $\mathbf{b}_{\text{PRS}} \in \{0, 1\}^{tb}$ be the known PRS bitstream, where $tb$ is the total number of bits. Given a secret key $key$ and a unique nonce-counter pair $nonce\parallel counter$ (where $\parallel$ denotes concatenation), AES-CTR generates a keystream $\mathbf{p} \in \{0, 1\}^{tb}$ by encrypting successive counter blocks of the form $nonce \parallel (counter{+i})$ using $key$. Here, the $\mathit{nonce}$ serves as a non-repeating initialization value to ensure that each encryption instance produces a unique keystream. Finally, the encrypted bitstream $\mathbf{b}_{\text{PRSenc}}$ is obtained as $\mathbf{b}_{\text{PRSenc}} = \mathbf{b}_{\text{PRS}} \oplus \mathbf{p}$.
The resulting encrypted bitstream $\mathbf{b}_{\text{PRSenc}}$ is then used to generate a complex vector $\mathbf{s}_{\text{PRSenc}} \in \mathbb{C}^{N}$ by mapping groups of bits to symbols from a chosen constellation (e.g., QPSK, where $tb=2N$). The original, unencrypted PRS sequence $\mathbf{s}_{\text{PRS}} \in \mathbb{C}^{N}$ is generated from $\mathbf{b}_{\text{PRS}}$ using the same constellation mapping.

To ensure signal integrity, $\mathbf{s}_{\text{PRSenc}}$ must satisfy three properties:
\begin{enumerate}[label=(\roman*)]
    \item For any receiver lacking the secret key, it must be statistically indistinguishable from noise, preventing unauthorized entities from detecting the signal via standard correlation. This randomness minimizes cross-correlation with legacy $\mathbf{s}_{\text{PRS}}$, preventing valid legacy transmissions from interfering with secure detection, while rendering unencrypted replicas ineffective for spoofing.
    \item It must exhibit strong autocorrelation to yield distinct peaks for legitimate receivers, to support OTDOA positioning.
    \item Sequences generated with different keys must exhibit minimal cross-correlation, ensuring that the target replica isolates the desired BS peak from frequency-multiplexed neighbors.
\end{enumerate}
These properties are detailed below.

\subsubsection{Cross-Correlation with Unencrypted PRS}

The cross-correlation $R$ of the deterministic, unencrypted $\mathbf{s}_{\text{PRS}}$ with encrypted, pseudorandom $\mathbf{s}_{\text{PRSenc}}$ sequences
\begin{equation}
R = \mathbf{s}_{\text{PRS}}^H \mathbf{s}_{\text{PRSenc}} = \sum\nolimits_{n=1}^N s_{\text{PRS},n}^* s_{\text{PRSenc},n}.
\end{equation}
is used to
assess
three scenarios: (i) detection attempts by an attacker using $\mathbf{s}_{\text{PRS}}$ as a template, (ii) spoofing attempts where a legitimate receiver correlates a local encrypted reference against an unencrypted attack signal, and (iii) interference from legitimate BSs that uses standard PRS.

The expectation of $R$ is 
\begin{equation}
    \mathbb{E}[R] = \mathbf{s}_{\text{PRS}}^H \mathbb{E}[\mathbf{s}_{\text{PRSenc}}] = 0,
\end{equation}
given the zero-mean property of standard, centrally symmetric constellations (e.g., $M$-PSK, $M$-QAM). To quantify the spread of the correlation output, we examine its variance. Since $\mathbb{E}[R]=0$, the variance is 
\begin{equation}
\mathrm{Var}(R) = \mathbb{E}[|R|^2] = \mathbf{s}_{\text{PRS}}^H \mathbf{C}_{\text{PRSenc}} \mathbf{s}_{\text{PRS}},  
\end{equation}
 where $\mathbf{C}_{\text{PRSenc}}$ is the covariance matrix of the encrypted signal. Since AES-CTR output is computationally indistinguishable from a uniform random source~\cite{hellekalek2003empirical, 6849074}, we model samples $s_{\text{PRSenc},n}$ as i.i.d., simplifying the covariance to ${\mathbf{C}_{\text{PRSenc}} = \sigma_{s}^2 \mathbf{I}_N}$, where $\sigma_{s}^2$ is the average symbol power. Substituting this into the variance equation yields
\begin{equation}
\mathrm{Var}(R) = \sigma_{s}^2 \lVert\mathbf{s}_{\text{PRS}}\rVert_2^2 = N \cdot (\sigma_{s}^2)^2.\label{e:var}
\end{equation}
Eq.~\eqref{e:var} shows that the cross-correlation energy scales linearly with $N$, 
significantly lower than the quadratic scaling $N^2$ of a valid autocorrelation peak. As a result: (i) an attacker cannot detect the presence of $\mathbf{s}_{\text{PRSenc}}$ using standard unencrypted templates; (ii) a legitimate receiver will observe no significant correlation peak when targeted by an unencrypted spoofing signal; and (iii) legacy transmissions from non-encrypted BSs will not interfere with secure PRS detection.

\subsubsection{Correlation at Legitimate UEs}

Accurate OTDOA positioning requires signals with strong autocorrelation properties. We argue that AES-CTR encryption preserves this characteristic, as its keystream exhibits cryptographic white-noise behavior, a characteristic supported by the fundamental statistical analyses of the cipher~\cite{6849074}. When the PRS is XORed with this keystream, the resulting encrypted sequence inherits the statistical indistinguishability from random noise. This behavior is consistent across the varying sequence lengths imposed by 5G bandwidth configurations and is supported by established empirical evidence. Specifically, Hellekalek and Wegenkittl~\cite{hellekalek2003empirical} demonstrate that AES outputs are indistinguishable from a uniform source starting from the first block, ensuring the absence of initialization bias even for short sequences. Furthermore, Zhang and Gong~\cite{6849074} verified that AES satisfies Golomb’s autocorrelation property for sequences of up to $2^{23}$ bits.

\subsubsection{Cross-Correlation of Distinct Encrypted PRSs}
Sequences generated with distinct keys must exhibit minimal cross-correlation to isolate the target BS peak from frequency-multiplexed neighbors. AES-CTR encryption enforces this quasi-orthogonality due to the global avalanche characteristics of the cipher, whereby a one-bit change in the key or initialization vector produces a statistically independent output stream, regardless of whether diversity arises from different keys or PRS sequences~\cite{zhou2018cryptographic, singh2025distinguishing}. Consequently, the cross-correlation between distinct encrypted PRSs approaches that of independent white-noise processes and remains negligible.
\subsection{Authentication Schemes Embedded in Empty PRS REs}
\label{sec:sig_embedding}
\noindent To provide explicit source authentication at the PL, we adopt the embedding strategy of~\cite{spanos}, which utilizes the PRS comb structure to insert cryptographic tags in empty PRS REs. We enhance this scheme by incorporating two key extensions: the support for asymmetric PRS DS and the addition of a forward error correction (FEC) layer.

\begin{figure}[htbp]
    \centering
    \begin{subfigure}[b]{\linewidth}
        \centering
        \resizebox{0.75\linewidth}{!}{
            \begin{tikzpicture}[%
    x=8mm, y=8mm, 
    font=\sffamily,
    newCellStyle/.style={
        shading=radial,
        inner color=#1!30!white,
        outer color=#1!80!black,
        draw=gray!50!black,
        thin
    }
  ]

  \foreach \r in {1,...,12} {
    \pgfmathtruncatemacro\rprime{mod((\r-1),6)+1}

    \foreach \c in {1,...,14} {
      \ifnum\c>12
        \filldraw[fill=white, draw=gray!50!black, thin] (\c-1,\r-1) rectangle ++(1,1);
      \else
        \ifnum\rprime=1 
          \ifnum\c=1 \def\cellcol{PRS1}\else\ifnum\c=2 \def\cellcol{SIG1}\else\ifnum\c=3 \def\cellcol{SIG3}\else\ifnum\c=4 \def\cellcol{PRS3}\else\ifnum\c=5 \def\cellcol{SIG2}\else\ifnum\c=6 \def\cellcol{PRS2}\else\ifnum\c=7 \def\cellcol{PRS1}\else\ifnum\c=8 \def\cellcol{SIG1}\else\ifnum\c=9 \def\cellcol{SIG3}\else\ifnum\c=10 \def\cellcol{PRS3}\else\ifnum\c=11 \def\cellcol{SIG2}\else\ifnum\c=12 \def\cellcol{PRS2}\fi\fi\fi\fi\fi\fi\fi\fi\fi\fi\fi\fi
        \else\ifnum\rprime=2 
          \ifnum\c=1 \def\cellcol{PRS2}\else\ifnum\c=2 \def\cellcol{SIG2}\else\ifnum\c=3 \def\cellcol{PRS1}\else\ifnum\c=4 \def\cellcol{SIG1}\else\ifnum\c=5 \def\cellcol{SIG3}\else\ifnum\c=6 \def\cellcol{PRS3}\else\ifnum\c=7 \def\cellcol{PRS2}\else\ifnum\c=8 \def\cellcol{SIG2}\else\ifnum\c=9 \def\cellcol{PRS1}\else\ifnum\c=10 \def\cellcol{SIG1}\else\ifnum\c=11 \def\cellcol{SIG3}\else\ifnum\c=12 \def\cellcol{PRS3}\fi\fi\fi\fi\fi\fi\fi\fi\fi\fi\fi\fi
        \else\ifnum\rprime=3 
          \ifnum\c=1 \def\cellcol{PRS3}\else\ifnum\c=2 \def\cellcol{SIG3}\else\ifnum\c=3 \def\cellcol{PRS2}\else\ifnum\c=4 \def\cellcol{SIG2}\else\ifnum\c=5 \def\cellcol{PRS1}\else\ifnum\c=6 \def\cellcol{SIG1}\else\ifnum\c=7 \def\cellcol{PRS3}\else\ifnum\c=8 \def\cellcol{SIG3}\else\ifnum\c=9 \def\cellcol{PRS2}\else\ifnum\c=10 \def\cellcol{SIG2}\else\ifnum\c=11 \def\cellcol{PRS1}\else\ifnum\c=12 \def\cellcol{SIG1}\fi\fi\fi\fi\fi\fi\fi\fi\fi\fi\fi\fi
        \else\ifnum\rprime=4 
          \ifnum\c=1 \def\cellcol{SIG1}\else\ifnum\c=2 \def\cellcol{PRS1}\else\ifnum\c=3 \def\cellcol{PRS3}\else\ifnum\c=4 \def\cellcol{SIG3}\else\ifnum\c=5 \def\cellcol{PRS2}\else\ifnum\c=6 \def\cellcol{SIG2}\else\ifnum\c=7 \def\cellcol{SIG1}\else\ifnum\c=8 \def\cellcol{PRS1}\else\ifnum\c=9 \def\cellcol{PRS3}\else\ifnum\c=10 \def\cellcol{SIG3}\else\ifnum\c=11 \def\cellcol{PRS2}\else\ifnum\c=12 \def\cellcol{SIG2}\fi\fi\fi\fi\fi\fi\fi\fi\fi\fi\fi\fi
        \else\ifnum\rprime=5 
          \ifnum\c=1 \def\cellcol{SIG2}\else\ifnum\c=2 \def\cellcol{PRS2}\else\ifnum\c=3 \def\cellcol{SIG1}\else\ifnum\c=4 \def\cellcol{PRS1}\else\ifnum\c=5 \def\cellcol{PRS3}\else\ifnum\c=6 \def\cellcol{SIG3}\else\ifnum\c=7 \def\cellcol{SIG2}\else\ifnum\c=8 \def\cellcol{PRS2}\else\ifnum\c=9 \def\cellcol{SIG1}\else\ifnum\c=10 \def\cellcol{PRS1}\else\ifnum\c=11 \def\cellcol{PRS3}\else\ifnum\c=12 \def\cellcol{SIG3}\fi\fi\fi\fi\fi\fi\fi\fi\fi\fi\fi\fi
        \else\ifnum\rprime=6 
          \ifnum\c=1 \def\cellcol{SIG3}\else\ifnum\c=2 \def\cellcol{PRS3}\else\ifnum\c=3 \def\cellcol{SIG2}\else\ifnum\c=4 \def\cellcol{PRS2}\else\ifnum\c=5 \def\cellcol{SIG1}\else\ifnum\c=6 \def\cellcol{PRS1}\else\ifnum\c=7 \def\cellcol{SIG3}\else\ifnum\c=8 \def\cellcol{PRS3}\else\ifnum\c=9 \def\cellcol{SIG2}\else\ifnum\c=10 \def\cellcol{PRS2}\else\ifnum\c=11 \def\cellcol{SIG1}\else\ifnum\c=12 \def\cellcol{PRS1}\fi\fi\fi\fi\fi\fi\fi\fi\fi\fi\fi\fi
        \fi\fi\fi\fi\fi\fi 

        \path[newCellStyle={\cellcol}] (\c-1,\r-1) rectangle ++(1,1);
      \fi
    }
  }

  \foreach \r in {1,...,12} {
    \node[left=2mm,font=\Large\sffamily] at (0,\r-1+0.5) {\r};
  }
  \node[rotate=90,font=\LARGE\bfseries\sffamily] at (-1.5,6) {Subcarriers};

  \foreach \c in {1,...,14} {
    \node[below=2mm,font=\Large\sffamily] at (\c-1+0.5,0) {\c};
  }
  
  \node[below=10mm,font=\LARGE\bfseries\sffamily] (xtitle) at (6.5,0) {OFDM Symbols};

  \matrix[
    matrix of nodes,
    below=3mm of xtitle, 
    nodes={
        minimum width=2.8cm, 
        minimum height=6mm, 
        font=\bfseries\sffamily
    },
    column sep=2mm, 
    row sep=2mm
  ] {
    |[newCellStyle={PRS1}]| PRS1  &
    |[newCellStyle={PRS2}]| PRS2  &
    |[newCellStyle={PRS3}]| PRS3  \\
    |[newCellStyle={SIG1}, text=\sigtxtcol]| SIG1  &
    |[newCellStyle={SIG2}, text=\sigtxtcol]| SIG2  &
    |[newCellStyle={SIG3}, text=\sigtxtcol]| SIG3  \\
  };

\end{tikzpicture}%
        } 
        \caption{Comb Size 6}
    \end{subfigure}\\[1ex] 
    
    \begin{subfigure}[b]{\linewidth}
        \centering
        \resizebox{0.75\linewidth}{!}{%
            \input{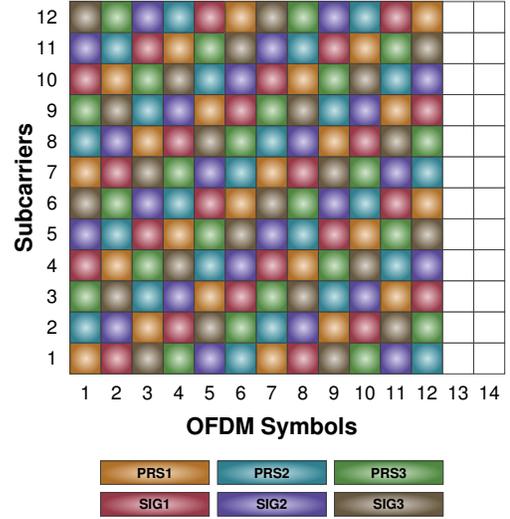}
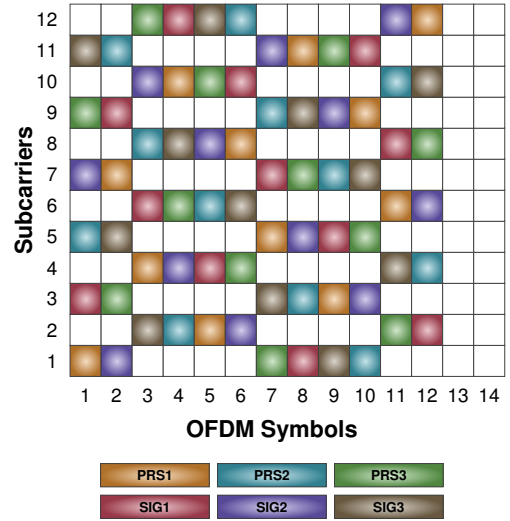%
        }
        \caption{Comb Size 12}
    \end{subfigure}
    
    \caption{PRS and signature dispersion for $K_{\text{comb}}^{\text{PRS}} \in \{6, 12\}$.}
    \label{PRSDispersion}
\end{figure}

The PRS utilizes a comb-like frequency structure where active subcarriers are spaced by $K_{\mathrm{comb}}^{\mathrm{PRS}} \in \{2, 4, 6, 12\}$ REs~\cite{3gppsrs}. While in standard operations the empty REs are reserved to allow neighboring BSs to transmit their PRS orthogonally by using distinct frequency offsets, the method in~\cite{spanos} repurposes a subset of these REs for authentication. In our implementation, the payload $\mathcal{T}$ can be configured as either a symmetric HMAC~\cite{bellare1996keying} or the proposed asymmetric DS~\cite{stinson2005cryptography}. To ensure reliability, the resulting bits are low-density parity-check (LDPC)-encoded before being modulated (e.g., QPSK) and mapped to the empty slot REs as per Fig.~\ref{PRSDispersion}. Upon reception, the UE extracts the symbols from these designated REs, performs LDPC decoding and demodulation, and validates the payload using a pre-shared key for HMAC or the public key of the BS for DS.

Deploying this mechanism introduces a trade-off between security and network capacity. Occupying these empty REs reduces the maximum number of frequency multiplexed BS transmissions. For instance, using a comb size of $K_{\mathrm{comb}}^{\mathrm{PRS}}=6$ normally supports up to six simultaneous orthogonal BS transmissions, multiplexed in frequency domain. However, embedding authentication data alongside the PRS typically requires halving this capacity (e.g., supporting only three orthogonal BSs) to accommodate the tag bandwidth~\cite{spanos}. Despite this reduction, the mechanism provides a robust defense against selective PRS spoofing attacks, as an adversary lacking the cryptographic key cannot generate a valid tag that matches the transmitted signal. Additionally, it aids in jamming detection, as interference corrupts the signature even if the PRS peak remains visible. However, this sensitivity introduces a secondary trade-off, as high noise levels may compromise signature decoding, potentially triggering false-alarms.

\subsection{Angular-Based Source Authentication (ABSA)}
\label{sec:absa}

\noindent While cryptographic techniques secure signal content, they do not verify transmitter location. To address this, we exploit spatial diversity using a uniform linear array (ULA) equipped at the UE. Assuming a known zenith angle (resolved \emph{a priori} or via alternative array configurations), we focus solely on azimuth estimation. The UE estimates the angle-of-arrival (AoA) $\theta_{\text{BS}}$ via super-resolution algorithms such as Multiple Signal Classification (MUSIC)~\cite{MUSIC} or Estimation of Signal Parameters via Rotational Invariance Techniques (ESPRIT)~\cite{ESPRIT}. This estimate is compared against a reference $\theta_{{\text{BS}_\text{ref}}}$, derived from prior beamforming reports or geometric calculations based on the last known user position. 

Spatial authentication verifies that the deviation remains within a predefined tolerance $\delta_{\text{th}}$, satisfying 
\begin{equation}
  \big|{\theta_{\text{BS}}} - \theta_{{\text{BS}_\text{ref}}}\big| \leq \delta_{\text{th}}.   
\end{equation}
This threshold accounts for estimation errors and multipath effects. A violation exceeding this bound flags a spatial anomaly, identifying the source as illegitimate. This technique robustly detects spoofing, jamming, and meaconing, as attack signals originate from the PL of the adversary rather than the legitimate BS. Consequently, the measured AoA significantly deviates from $\theta_{{\text{BS}_\text{ref}}}$, triggering an alarm.

\subsection{Downlink-Uplink Handshaking (DL-UL HS)}
\label{sec:handshaking}

\noindent Expanding the verification scope beyond the DL, we introduce a bidirectional consistency check. We propose a cross-layer verification protocol that explicitly links the DL and UL positioning procedures. This relies on the premise that adversaries cannot simultaneously manipulate both links simultaneously.
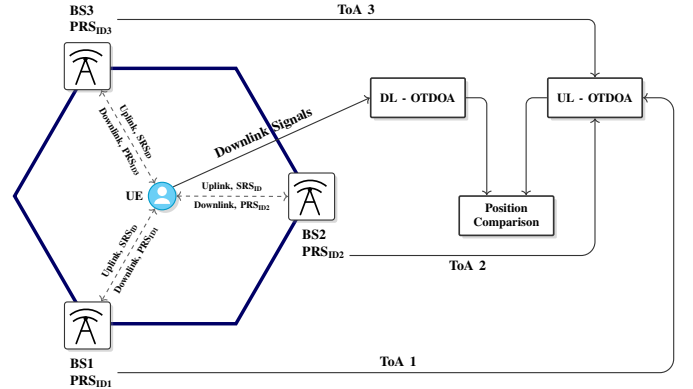
\begin{figure}[!t]
    \centering
    \resizebox{\linewidth}{!}{%
        \begin{tikzpicture}[
    font=\sffamily\scriptsize,
    bs_node/.style={
        rectangle,
        draw=black!80,
        thick,
        minimum size=1.1cm,
        align=center,
        fill=white,
        drop shadow={opacity=0.3, shadow xshift=2pt, shadow yshift=-2pt},
        rounded corners=2pt
    },
    ue_node/.style={
        circle,
        fill=cyan!50,
        draw=cyan!80!black,
        thick,
        minimum size=0.7cm,
        inner sep=0pt,
        label={[xshift=-0.7cm, yshift = -0.6cm, font=\bfseries\small]UE},
        drop shadow={opacity=0.3}
    },
    block/.style={
        rectangle,
        draw=black!80,
        very thick,
        fill=white,
        align=center,
        minimum width=2.4cm,
        minimum height=1.0cm,
        drop shadow={opacity=0.3},
        font=\bfseries\small,
        rounded corners=1pt
    }
]

\newcommand{\bsdrawing}{
    \tikz[baseline=-0.1cm, scale=0.8]{
        \draw[line width=1.2pt, line cap=round] (-0.3,-0.4) -- (0,0.4) -- (0.3,-0.4); 
        \draw[line width=1.2pt, line cap=round] (-0.15,0) -- (0.15,0); 
        \draw[fill=black] (0,0.4) circle (1.5pt); 
        \draw[line width=1.2pt] (-0.4,0.3) arc (130:50:0.6); 
        \draw[line width=1.2pt] (-0.5,0.5) arc (130:50:0.8); 
    }
}

\node[regular polygon, regular polygon sides=6, minimum size=7.5cm, draw=blue!40!black, line width=2.5pt, shape border rotate=0] (hex) {};

\node[ue_node] (ue) at (hex.center) {};
\begin{scope}[shift={(ue.center)}]
    \fill[white] (0,0.12) circle (0.12); 
    \fill[white] (-0.2,-0.2) to[out=90,in=180] (0,-0.05) to[out=0,in=90] (0.2,-0.2) -- cycle; 
\end{scope}

\node[bs_node, left=-0.6cm and -0.6cm of hex.corner 4] (bs1) {\bsdrawing};
\node[below=3pt, xshift=0.1cm, font=\bfseries, align=left] (bs1_label) at (bs1.south) {BS1\\PRS\textsubscript{ID1}};

\node[bs_node, above right=-0.6cm and -0.6cm of hex.corner 2] (bs3) {\bsdrawing};
\node[above=20pt, xshift=-0.5cm, font=\bfseries, align=left] (bs3_label) at (bs3.east) {BS3\\PRS\textsubscript{ID3}};

\node[bs_node, below right=-0.6cm and -0.6cm of hex.corner 6] (bs2) {\bsdrawing};
\node[below=3pt, xshift=0.3cm, font=\bfseries, align=left] (bs2_label) at (bs2.south) {BS2\\PRS\textsubscript{ID2}};

\node[block] (ul_otdoa) at (11.0, 2.5) {UL - OTDOA};
\node[block] (dl_otdoa) at (6.5, 2.5) {DL - OTDOA};
\node[block] (comp) at (8.75, -0.5) {Position\\Comparison};

\draw[<->, thick, dashed, draw=black!60] (ue) -- 
    node[sloped, above, font=\bfseries\scriptsize] {Uplink, SRS\textsubscript{ID}} 
    node[sloped, below, font=\bfseries\scriptsize] {Downlink, PRS\textsubscript{ID1}} 
    (bs1);

\draw[<->, thick, dashed, draw=black!60] (ue) -- 
    node[sloped, above, font=\bfseries\scriptsize] {Uplink, SRS\textsubscript{ID}} 
    node[sloped, below, font=\bfseries\scriptsize] {Downlink, PRS\textsubscript{ID2}} 
    (bs2);

\draw[<->, thick, dashed, draw=black!60] (ue) -- 
    node[sloped, above, font=\bfseries\scriptsize] {Uplink, SRS\textsubscript{ID}} 
    node[sloped, below, font=\bfseries\scriptsize] {Downlink, PRS\textsubscript{ID3}} 
    (bs3);

\draw[->, thick, rounded corners=10pt, draw=black!80] 
    (bs1_label.east) 
    -- (13.0, 0 |- bs1_label.east) coordinate(far_right) 
    -- node[pos=0.2, above, font=\bfseries, xshift=-7cm, yshift=-1.37cm] {ToA 1} (far_right |- ul_otdoa.east)
    -- (ul_otdoa.east);

\draw[->, thick, rounded corners=5pt, draw=black!80] 
    (bs3_label.east) -| node[pos=0.25, above, font=\bfseries] {ToA 3} (ul_otdoa.north);

\draw[->, thick, rounded corners=10pt, draw=black!80] 
    ($(bs2_label.east) + (0, -0.3)$) 
    -- node[pos=2, below, font=\bfseries] {ToA 2} ++(1.5, 0) coordinate(bs2_path)
    -- (bs2_path -| ul_otdoa.south)
    -- (ul_otdoa.south);

\draw[->, thick, draw=black!80] 
    (ue.40) -- node[pos=0.485, sloped, above, font=\bfseries] {Downlink Signals} (dl_otdoa.west);

\draw[->, thick, rounded corners=5pt, draw=black!80] 
    (ul_otdoa.west) -| ($(comp.north) + (0.5, 0)$);

\draw[->, thick, rounded corners=5pt, draw=black!80] 
    (dl_otdoa.east) -| ($(comp.north) + (-0.5, 0)$);

\end{tikzpicture}%
    }
    \caption{Downlink-uplink handshaking protocol procedure.}
    \label{handshaking}
\end{figure}
As illustrated in Fig.~\ref{handshaking}, the UE computes its position $\widehat{\mathbf{p}}_{\text{UE,DL}}$ via DL-OTDOA and securely embeds the result in its uplink transmission (e.g., via the physical uplink shared channel (PUSCH) or the sounding reference signal (SRS) empty REs). When the network of serving BSs receives the UE uplink signals, it performs an independent UL-OTDOA estimation to determine the UE position, $\widehat{\mathbf{p}}_{\text{UE,UL}}$.

The network core or the location management function (LMF) validates consistency by verifying 
\begin{equation}
\big\|\widehat{\mathbf{p}}_{\text{UE,DL}} - \widehat{\mathbf{p}}_{\text{UE,UL}}\big\| \leq \epsilon.    
\end{equation}
A discrepancy that exceeds the threshold $\epsilon$ triggers an integrity alarm. This mechanism robustly counters spoofing, replay, and jamming, as DL manipulation forces the UE to calculate a falsified $\widehat{\mathbf{p}}_{\text{UE,DL}}$. Assuming that the attacker cannot alter physical UL propagation to match this falsified location, the estimates diverge, in this way exposing the attack.

\subsection{Position Tracking}

\noindent The final mechanism exploits the temporal continuity of the UE trajectory to detect anomalies using a recursive Bayesian estimator based on a typical KF~\cite{welch1995introduction} to track the UE position and velocity over time. The UE motion is modeled with a constant-velocity dynamic model, where the state vector $\mathbf{x_k} = [p_x,\, p_y,\, v_x,\, v_y]^T$ represents 2-D position and velocity. State evolution follows
\begin{equation}
\mathbf{x}_{k+1} = \mathbf{F}\mathbf{x}_k + \mathbf{w}_k, \quad \mathbf{w}_k \sim \mathcal{N}(\mathbf{0}, \mathbf{Q}),
\end{equation}
with $\mathbf{F}$ denoting the state transition matrix and $\mathbf{Q}$ the process noise covariance matrix, constructed using the discrete \mbox{white-noise} acceleration model for random maneuvers~\cite{bar2001estimation}.

Attack detection is performed using innovation-based statistical gating~\cite{mehra1971innovations,willsky1974adaptive}. The innovation vector ${\tilde{\mathbf{y}}_k=\mathbf{z}-\mathbf{H} \hat{\mathbf{x}}_{k|k-1}}$ quantifies the discrepancy between the incoming 5G measurement $\mathbf{z}_k$ and the predicted state. Under nominal conditions, the normalized innovation squared (NIS), denoted as
\begin{equation}
  \epsilon_k=\tilde{\mathbf{y}}_k^T \mathbf{S}_k^{-1} \tilde{\mathbf{y}}_k,  
\end{equation}
where $\mathbf{S}_k^{-1}$ is the innovation covariance, follows a Chi-square distribution with two degrees of freedom. An attack is flagged if $\epsilon_k$ exceeds a dynamic threshold $\gamma$ or if the measurement is invalid. Upon detection, the measurement update is skipped, and the state is propagated using prediction only, while the error covariance $\mathbf{P}$ is allowed to grow to reflect increasing uncertainty.

A known vulnerability of recursive estimators in adversarial settings is that, during extended coasting periods, the repeated accumulation of $\mathbf{Q}$ inflates the predicted error covariance $\mathbf{P}_{k|k-1}$ and the innovation covariance $\mathbf{S}_k$. As a result, even a highly erroneous measurement may fall within the inflated gate, enabling false state injection upon signal recovery. While prior work mitigates this effect by fusing external proprioceptive sensors~\cite{9868262}, we instead utilize a sequential validity test based on $M$-of-$N$ logic. Rather than accepting the first post-attack measurement, we require kinematic consistency over a window of $N$ consecutive epochs, enforcing temporal coherence without relying on external sensors.


\section{Simulation Framework}
\label{simulation_framework}
\noindent We evaluate the proposed techniques using VeriLoc, with numerology $\mu=0$, 10 MHz bandwidth, and 122.88 MHz sampling rate. The simulation considers the transmission of a standard radio frame comprising 10 slots, where PRS resources are mapped to every slot within the frame. Each PRS instance is configured with a comb size of $K_{\mathrm{comb}}^{\mathrm{PRS}}=6$, spanning 12 consecutive OFDM symbols. This high-density configuration is chosen to maximize the correlation gain, thereby optimizing positioning accuracy under nominal conditions and enhancing signal robustness against adversarial interference. Embedded authentication payloads use $1/2$ LDPC code rate with~25~decoding iterations. For ABSA, we consider a five-element ULA, with AoA estimated via ESPRIT.
\begin{figure}[t]
    \centering
    \begin{subfigure}[b]{1.0\linewidth}
        \centering
       \includegraphics[width=1\textwidth]{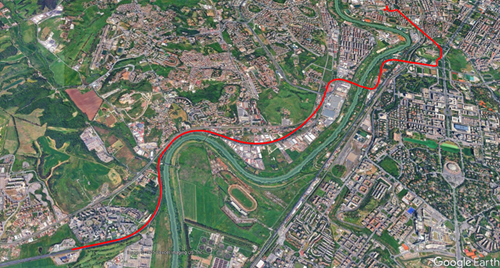}
        \caption{Observed trajectory depicted in Google Earth.}
        \label{traj_google}
    \end{subfigure}
    
    \begin{subfigure}[b]{\linewidth}
        \centering
       \includegraphics[width=1\textwidth]{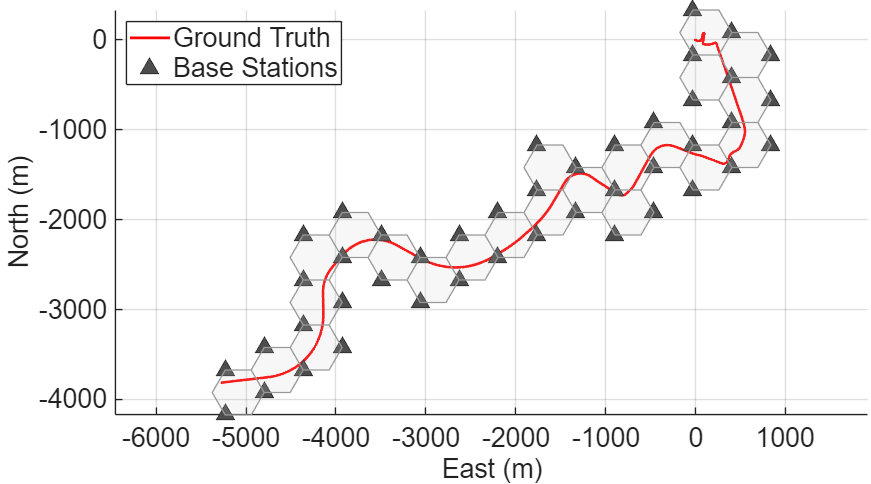}
        \caption{VeriLoc simulation model showing the trajectory.}
        \label{traj_matlab}
    \end{subfigure}
    
    \caption{Real-world trajectory and VeriLoc simulation model.}
    \label{trajectory_viz}
\end{figure}

The system operates at a carrier frequency of 3.5 GHz. The network topology assumes a hexagonal cellular layout with an inter-site distance (ISD) of 500 m~\cite{ISD}, corresponding to a cell side length of approximately 289 m. The topology is randomized in each iteration to ensure statistically unbiased results. The propagation environment models a dense urban canyon scenario, characterized by an average building height of 15 m and a street width of 5 m. The BS and UE antennas heights are configured at 25 m and 1.5 m, respectively~\cite{3gppheights}. The evaluation follows a dynamic~\mbox{1200-point} trajectory, depicted in Fig.~\ref{trajectory_viz}, with 1 s resolution between consecutive location estimates, where a non-uniform velocity profile introduces a point-to-point variation in the Doppler shift.

To assess the robustness of the proposed techniques against PL threats, the system was evaluated under the three adversarial models of Section~\ref{subsec:threat_models}: selective PRS spoofing, meaconing, and jamming. The simulation timeline is partitioned into three phases to isolate the impact of the attacks:
\begin{enumerate}
    \item \textbf{Initial Benign Phase (Points 1–300):} The system operates under nominal conditions with no adversary present.
    \item \textbf{Attack Phase (Points 301–900):} The adversary actively interferes with the legitimate signal.
    \item \textbf{Recovery Phase (Points 901–1200):} The system returns to benign conditions.
\end{enumerate}

In all adversarial scenarios, the attacker is modeled as a mobile node that closely tracks the target UE, following the UE exact trajectory  with a fixed delay of 10 trajectory points (10 s lag).

\section{Results and Discussion}
\label{results}
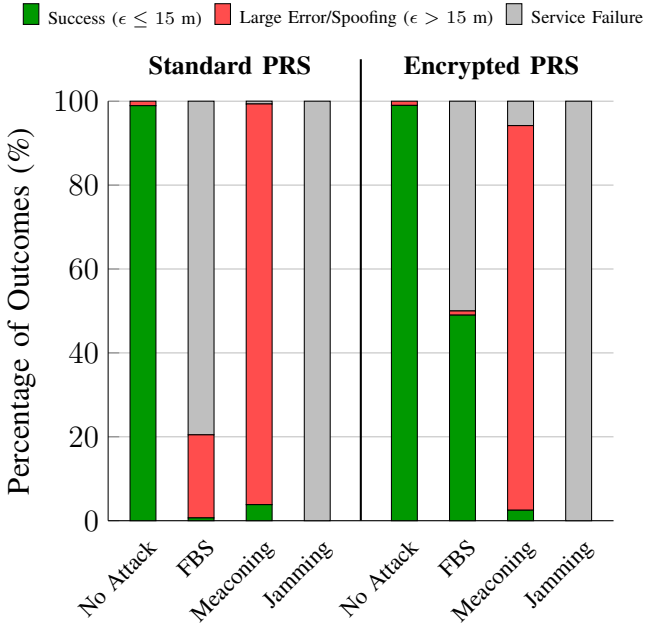
\begin{figure}[htbp]
    \centering
    \resizebox{\linewidth}{!}{%
        \begin{tikzpicture}
    \begin{axis}[
        ybar stacked,
        bar width=10pt,
        ymin=0, ymax=100,
        ylabel={Percentage of Outcomes (\%)},
        clip=false,
        xtick={1, 2, 3, 4, 5.5, 6.5, 7.5, 8.5},
        xticklabels={No Attack, FBS, Meaconing, Jamming, No Attack, FBS, Meaconing, Jamming},
        xticklabel style={font=\small, rotate=45, anchor=north east, align=right, yshift=-2pt},
        yticklabel style={font=\large},
        ylabel style={font=\large},
        legend style={at={(0.45,1.15)}, anchor=south, legend columns=-1, font=\scriptsize, draw=none, /tikz/every even column/.append style={column sep=0.1cm}, /tikz/every odd column/.append style={column sep=0.05cm}},
        axis lines*=left,
        enlarge x limits={abs=0.6},
        grid=major,
    ]
    \addplot+[fill=green!60!black, draw=black] coordinates {(1, 98.93) (2, 0.67) (3, 3.83) (4, 0) (5.5, 99) (6.5, 49) (7.5, 2.5) (8.5, 0)};
    \addplot+[fill=red!70!white, draw=black] coordinates {(1, 1.07) (2, 19.8) (3, 95.51) (4, 0) (5.5, 1) (6.5, 1) (7.5, 91.68) (8.5, 0)};
    \addplot+[fill=gray!50, draw=black] coordinates {(1, 0) (2, 79.53) (3, 0.67) (4, 100) (5.5, 0) (6.5, 50) (7.5, 5.82) (8.5, 100)};
    \legend{Success ($\epsilon \le 15$ m), Large Error/Spoofing ($\epsilon > 15$ m), Service Failure}
    \draw [black, solid, thick] (axis cs:4.75,0) -- (axis cs:4.75,110);
    \node[font=\bfseries\normalsize, anchor=north] at (axis cs:2.5, 113) {Standard PRS};
    \node[font=\bfseries\normalsize, anchor=north] at (axis cs:7.0, 113) {Encrypted PRS};
    \end{axis}
\end{tikzpicture}%
    }
    \caption{Performance comparison of encrypted PRS \emph{vs.} standard PRS signal under different conditions.}
    \label{encrypted_PRS}
\end{figure}

\noindent First, we assess the efficacy of the encrypted PRS relative to the standard PRS. Fig.~\ref{encrypted_PRS} details the comparative positioning performance, categorized into three distinct outcomes: accurate positioning (error {$\le$} 15 m), large positioning error (exceeding 15 m, indicating spoofing), and multilateration solver failure resulting in a DoS. Because this 15 m threshold corresponds directly to the 99.9th percentile of the cumulative distribution function (CDF) of the position estimation error under benign conditions, it serves as a strict baseline. Evaluated against this baseline, the encrypted PRS exhibits performance strictly comparable to the standard PRS, maintaining a high success rate. This confirms that the proposed encrypted PRS is fully compatible with standard OTDOA positioning operations in the absence of attacks.

Under an FBS attack, a distinct contrast in performance is observed. While the standard PRS remains highly susceptible to spoofing, the encrypted PRS effectively eliminates false positioning, exhibiting a strictly binary performance outcome: either a valid solution or a service failure (DoS), with no large errors observed. However, a DoS rate of approximately 50\% persists. This arises because the attacker transmits a high-power standard PRS on the same subcarriers as the encrypted signal. Although the receiver rejects the adversarial waveform as uncorrelated noise, the high-energy injection on the active resources significantly degrades the SINR of legitimate signals. Consequently, the spoofer effectively functions as a jammer, compromising the hearability of weak serving BSs, which mainly occurs when the UE is located near the cell edge. 

Finally, results confirm the theoretical limitations of PL encryption. As expected, encrypted PRS offers no inherent advantage against meaconing or jamming. This limitation stems from the fundamental nature of these attacks, as meaconing simply replays the intercepted signal without requiring the encryption key, while jamming injects wideband noise to degrade reception quality regardless of signal content.

\begin{figure}[b!]
    \centering
    \resizebox{\linewidth}{!}{%
        \begin{tikzpicture}
    \begin{axis}[
        ybar,
        bar width=3pt,
        ymin=0, ymax=100,
        ylabel={Correct Decision Rate (\%)},
        clip=false,
        symbolic x coords={No Attack, FBS, Meaconing, Jamming},
        xtick=data,
        xticklabel style={font=\small, align=center, yshift=-2pt},
        yticklabel style={font=\normalsize},
        ylabel style={font=\normalsize},
        ymajorgrids=true,
        grid style={dashed, gray!30},
        axis on top,
        axis line style={draw=none},
        axis x line*=bottom,
        axis y line*=left,
        enlarge x limits=0.2,
        legend style={
            at={(0.42,1.03)}, 
            anchor=south, 
            legend columns=5,  
            font=\scriptsize, 
            draw=none, 
            fill=none, 
            /tikz/every even column/.append style={column sep=0.15cm}, 
            /tikz/every odd column/.append style={column sep=0.08cm}
        }
    ]
    \addplot[fill=blue!60!black, draw=none, area legend] coordinates {(No Attack, 99.16) (FBS, 99.5) (Meaconing, 31.83) (Jamming, 100)};
    \addplot[fill=red!70!black, draw=none, area legend] coordinates {(No Attack, 98.75) (FBS, 100) (Meaconing, 31.67) (Jamming, 100)};
    \addplot[fill=teal!70!white, draw=none, area legend] coordinates {(No Attack, 96.77) (FBS, 99.37) (Meaconing, 99.83) (Jamming, 100)};
    \addplot[fill=orange!80!white, draw=none, area legend] coordinates {(No Attack, 99.39) (FBS, 99) (Meaconing, 92.67) (Jamming, 100)};
    \addplot[fill=gray!60!black, draw=none, area legend] coordinates {(No Attack, 95.72) (FBS, 98.33) (Meaconing, 39) (Jamming, 100)};
    \legend{HMAC, DS, ABSA, HS, Tracking}
    \end{axis}
\end{tikzpicture}%
    }
    \caption{Security evaluation under benign (false-alarm) and adversarial (threat detection) conditions.}
    \label{techniques}
\end{figure}
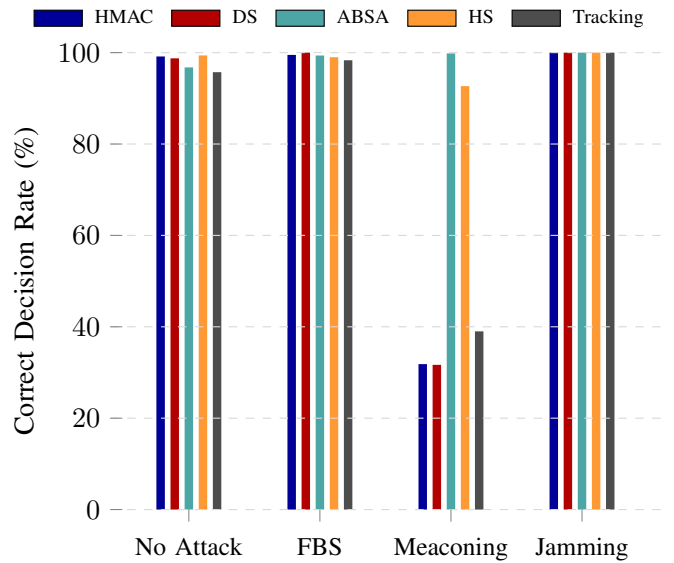

\begin{figure}[t!]
    \centering
    \begin{subfigure}[b]{\linewidth}
        \centering
        \resizebox{\linewidth}{!}{%
\pgfplotsset{
    attack zone/.style={
        extra x ticks={301, 900},
        extra x tick labels={},
        extra x tick style={grid=major, major grid style={red, dashed, thick}},
        execute at begin axis={
            \fill[red, opacity=0.15] (axis cs:301, -0.2) rectangle (axis cs:900, 1.2);
        }
    },
    logic plot style/.style={
        width=14cm, height=3.5cm,
        xmin=0, xmax=1200,
        ymin=-0.2, ymax=1.2,
        ytick={0,1},
        yticklabels={\textbf{Invalid},\textbf{Valid}},
        yticklabel style={font=\large\sffamily, align=right}, 
        xtick style={draw=none},
        xticklabels=\empty,
        enlarge x limits=false,
        const plot,
        no markers,
        axis on top,
        axis line style={black!60},
        grid=none
    }
}
\begin{tikzpicture}
    \begin{groupplot}[
        group style={
            group size=1 by 5,
            vertical sep=0.15cm,
            x descriptions at=edge bottom
        }
    ]

    \nextgroupplot[logic plot style, attack zone, ylabel={\large HMAC}]
    \addplot[blue!60!black, ultra thick] table [x=Time, y=HMAC, col sep=comma] {fbs_results.csv};

    \nextgroupplot[logic plot style, attack zone, ylabel={\large DS}]
    \addplot[red!70!black, ultra thick] table [x=Time, y=DigSig, col sep=comma] {fbs_results.csv};

    \nextgroupplot[logic plot style, attack zone, ylabel={\large ABSA}]
    \addplot[teal!70!white, ultra thick] table [x=Time, y=ABSA, col sep=comma] {fbs_results.csv};

    \nextgroupplot[logic plot style, attack zone, ylabel={\large HS}]
    \addplot[orange!80!white, ultra thick] table [x=Time, y=Handshake, col sep=comma] {fbs_results.csv};

    \nextgroupplot[
        logic plot style, 
        attack zone,
        ylabel={\large Tracking},
        xlabel={\Large Trajectory Points}, 
        xtick style={draw=black},       
        xticklabels={0, 200, 400, 600, 800, 1000, 1200},
        xtick={0, 200, 400, 600, 800, 1000, 1200},
        xticklabel style={font=\Large}
    ]
    \addplot[gray!60!black, ultra thick] table [x=Time, y=Tracking, col sep=comma] {fbs_results.csv};

    \end{groupplot}
\end{tikzpicture}%
        } 
        \caption{Selective PRS spoofing (FBS) attack scenario}
        \label{fig:fbs_ts}
    \end{subfigure}\vspace{1ex} 
    
    \begin{subfigure}[b]{\linewidth}
        \centering
        \resizebox{\linewidth}{!}{%
\pgfplotsset{
    attack zone/.style={
        extra x ticks={301, 900},
        extra x tick labels={},
        extra x tick style={grid=major, major grid style={red, dashed, thick}},
        execute at begin axis={
            \fill[red, opacity=0.15] (axis cs:301, -0.2) rectangle (axis cs:900, 1.2);
        }
    },
    logic plot style/.style={
        width=14cm, height=3.5cm,
        xmin=0, xmax=1200,
        ymin=-0.2, ymax=1.2,
        ytick={0,1},
        yticklabels={\textbf{Invalid},\textbf{Valid}},
        yticklabel style={font=\large\sffamily, align=right}, 
        xtick style={draw=none},
        xticklabels=\empty,
        enlarge x limits=false,
        const plot,
        no markers,
        axis on top,
        axis line style={black!60},
        grid=none
    }
}

\begin{tikzpicture}
    \begin{groupplot}[
        group style={
            group size=1 by 5,
            vertical sep=0.15cm,
            x descriptions at=edge bottom
        }
    ]

    \nextgroupplot[logic plot style, attack zone, ylabel={\large HMAC}]
    \addplot[blue!60!black, ultra thick] table [x=Time, y=HMAC, col sep=comma] {meaconing_results.csv};

    \nextgroupplot[logic plot style, attack zone, ylabel={\large DS}]
    \addplot[red!70!black, ultra thick] table [x=Time, y=DigSig, col sep=comma] {meaconing_results.csv};

    \nextgroupplot[logic plot style, attack zone, ylabel={\large ABSA}]
    \addplot[teal!70!white, ultra thick] table [x=Time, y=ABSA, col sep=comma] {meaconing_results.csv};

    \nextgroupplot[logic plot style, attack zone, ylabel={\large HS}]
    \addplot[orange!80!white, ultra thick] table [x=Time, y=Handshake, col sep=comma] {meaconing_results.csv};

    \nextgroupplot[
        logic plot style, 
        attack zone,
        ylabel={\large Tracking},
        xlabel={\Large Trajectory Points}, 
        xtick style={draw=black},       
        xticklabels={0, 200, 400, 600, 800, 1000, 1200},
        xtick={0, 200, 400, 600, 800, 1000, 1200},
        xticklabel style={font=\Large}
    ]
    \addplot[gray!60!black, ultra thick] table [x=Time, y=Tracking, col sep=comma] {meaconing_results.csv};

    \end{groupplot}
\end{tikzpicture}%
        }
        \caption{Meaconing attack scenario}
        \label{fig:meac_ts}
    \end{subfigure}\vspace{1ex}
    
    \begin{subfigure}[b]{\linewidth}
        \centering
        \resizebox{\linewidth}{!}{%
\pgfplotsset{
    attack zone/.style={
        extra x ticks={301, 900},
        extra x tick labels={},
        extra x tick style={grid=major, major grid style={red, dashed, thick}},
        execute at begin axis={
            \fill[red, opacity=0.15] (axis cs:301, -0.2) rectangle (axis cs:900, 1.2);
        }
    },
    logic plot style/.style={
        width=14cm, height=3.5cm,
        xmin=0, xmax=1200,
        ymin=-0.2, ymax=1.2,
        ytick={0,1},
        yticklabels={\textbf{Invalid},\textbf{Valid}},
        yticklabel style={font=\large\sffamily, align=right}, 
        xtick style={draw=none},
        xticklabels=\empty,
        enlarge x limits=false,
        const plot,
        no markers,
        axis on top,
        axis line style={black!60},
        grid=none
    }
}

\begin{tikzpicture}
    \begin{groupplot}[
        group style={
            group size=1 by 5,
            vertical sep=0.15cm,
            x descriptions at=edge bottom
        }
    ]

    \nextgroupplot[logic plot style, attack zone, ylabel={\large HMAC}]
    \addplot[blue!60!black, ultra thick] table [x=Time, y=HMAC, col sep=comma] {jamming_results.csv};

    \nextgroupplot[logic plot style, attack zone, ylabel={\large DS}]
    \addplot[red!70!black, ultra thick] table [x=Time, y=DigSig, col sep=comma] {jamming_results.csv};

    \nextgroupplot[logic plot style, attack zone, ylabel={\large ABSA}]
    \addplot[teal!70!white, ultra thick] table [x=Time, y=ABSA, col sep=comma] {jamming_results.csv};

    \nextgroupplot[logic plot style, attack zone, ylabel={\large HS}]
    \addplot[orange!80!white, ultra thick] table [x=Time, y=Handshake, col sep=comma] {jamming_results.csv};

    \nextgroupplot[
        logic plot style, 
        attack zone,
        ylabel={\large Tracking},
        xlabel={\Large Trajectory Points}, 
        xtick style={draw=black},       
        xticklabels={0, 200, 400, 600, 800, 1000, 1200},
        xtick={0, 200, 400, 600, 800, 1000, 1200},
        xticklabel style={font=\Large}
    ]
    \addplot[gray!60!black, ultra thick] table [x=Time, y=Tracking, col sep=comma] {jamming_results.csv};

    \end{groupplot}
\end{tikzpicture}%
        }
        \caption{Jamming attack scenario}
        \label{fig:jam_ts}
    \end{subfigure}

    \caption{Time-series detection performance under different attack vectors, with the active attack interval shaded in red.}
    \label{fig:all_attacks}
\end{figure}
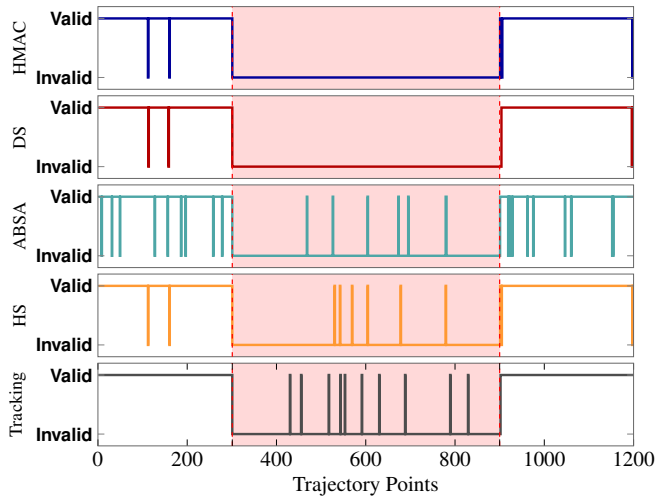
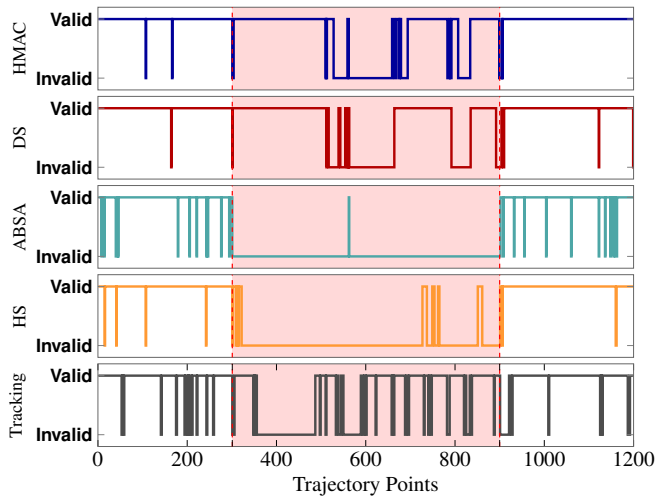
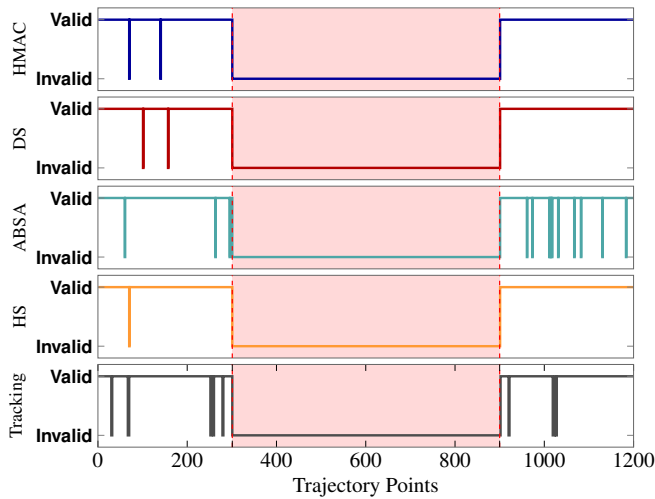

Following the encryption analysis, we assess the performance of the proposed threat detection techniques: HMAC/DS embedding, ABSA, DL-UL HS, and tracking. Fig.~\ref{techniques} summarizes the correct decision rates. This evaluation distinguishes between two key operational behaviors. Under benign conditions, it quantifies the robustness against false alarms, ensuring that security checks do not disrupt legitimate positioning services. Under adversarial conditions, it measures the attack detection capability of each technique. To ensure robustness against benign AoA estimation errors, we configure a 20\degree{} angular tolerance for ABSA, a 20 m discrepancy limit for HS, and an 85\% confidence interval $\gamma$, with a two-epoch window for tracking. These tolerance values were empirically resolved under benign, high SNR conditions, incorporating an additional margin to guarantee operational robustness. Fig.~\ref{fig:all_attacks} details the time-series response against the defined attack vectors.

Regarding false-alarm assessment, we exclude DoS instances caused by low signal quality, focusing strictly on cases where a positioning solution is available but incorrectly flagged as malicious. Under nominal conditions, the cryptographic techniques (HMAC/DS) and the DL-UL HS demonstrate near-perfect robustness. ABSA exhibits a slightly higher false-alarm rate, as the 20\degree{} threshold is sensitive to measurement errors at specific locations, (i.e. the edges of the cell), similar to observations for tracking and the $\gamma$ threshold. Finally, although signature-based methods are theoretically susceptible to false-alarms under low SNR conditions, where bit errors can corrupt the authentication tag, the results indicate they remain stable in the evaluated environment.

Performance under attack varies depending on the specific threat. Against FBS, all techniques maintain high detection rates. However, meaconing exposes vulnerabilities inherent in signature-based methods. Since meaconing physically replays the waveform, embedded cryptographic signatures remain valid upon retransmission. The observed detection rate (approx. 40\%) is not due to protocol security, but rather signal degradation; the accumulation of channel impairments of the links from the BS to the attacker and from the attacker to the UE often distorts the signature enough to cause verification failure. Similarly, tracking performance is limited to approx. 40\%, as the injected measurements closely resemble genuine data, yielding low NIS values that complicate detection.

In contrast, ABSA achieves complete meaconing detection by identifying angular discrepancies from a single source. Beyond detection, it resolves the malicious AoA, enabling localization of the threat direction. The DL-UL HS is similarly robust, exceeding a 90\% detection rate. Detection failures occur only when the adversary is in close proximity to the UE, where the meaconed signal yields a position estimate within the 20 m tolerance of the valid UL solution, thereby bypassing the integrity check and resulting in false authentication.

Finally, under wideband jamming, all techniques register an 100\% correct decision rate, albeit via different mechanisms. For HMAC, DS, and DL-UL HS, jamming interference prevents signature demodulation or DL position computation, invalidating the position estimate. ABSA, on the contrary, actively detects the threat by resolving the AoA of the jammer, while tracking detects the anomaly and enters coasting mode. Notably, while evaluated on unencrypted PRS, these techniques rely on signal geometry and data payload, ensuring effectiveness independent of PL encryption.

\section{Conclusion}
\label{conclusion}
\noindent This work presented a PL security framework designed to mitigate the vulnerabilities inherent in 5G OTDOA positioning. A foundational contribution of this research is the development and open-source release of VeriLoc, a system-level simulator engineered specifically for realistic PL threat injection and vulnerability assessment. Leveraging this platform, we proposed three distinct security enhancement techniques, namely PRS encryption based on AES-CTR, ABSA and DL-UL HS, alongside the implementation and assessment of established tracking and embedded authentication mechanisms, the latter extended in this work to support DSs.

Our analysis confirms that the proposed PRS encryption based on AES-CTR significantly improves resilience against selective PRS spoofing by preventing malicious signal generation, although it remains vulnerable to meaconing and DoS via high-power jamming. Furthermore, the evaluation highlights clear performance trade-offs among detection strategies. Signature-based schemes provide strong source integrity verification under FBS and jamming attacks; however, they are susceptible to false position validation in meaconing scenarios, where adversaries can successfully replay valid cryptographic tags. In contrast, ABSA and DL-UL HS exhibit the highest reliability. By exploiting spatial inconsistencies and positional discrepancies respectively, these techniques detect all considered threats, offering a geometric verification layer that operates independently of cryptographic protection. Overall, the results emphasize the need for a multi-layer security paradigm combining cryptographic prevention with PL detection to achieve resilience against diverse attack vectors.

\section*{Software Availability}
To support reproducible research and enable further advancements in 5G positioning security, the VeriLoc framework developed and utilized in this study has been made publicly available. The underlying source code, system architecture modules, and associated documentation can be accessed via the Zenodo repository at \url{https://zenodo.org/records/19051224}. Furthermore, the actively maintained source code and future updates can be accessed via the project's GitHub repository at \url{https://github.com/loctio/VeriLoc}.

\section*{Acknowledgments}
\noindent This work has been performed under the scope of the Technological Enablers of Cellular Networks for PVT Assurance (PARTICLE) activity, (NAVISP-EL1-077), funded by the European Space Agency (ESA).
\balance
\bibliographystyle{IEEEtran}
\bibliography{IEEEabrv,navitec}
\end{document}